\documentclass{vldb}
\usepackage{graphicx}
\usepackage{balance}  % for  \balance command ON LAST PAGE  (only there!)

\usepackage{listings}
\usepackage{xcolor}
\usepackage{textcomp}

\usepackage{amsthm}
\theoremstyle{definition}
\newtheorem{definition}{Definition}
\newtheorem{theorem}{Theorem}
\usepackage{booktabs}
\usepackage{algorithm}
\usepackage{algorithmic}

\usepackage{hyperref}
\usepackage{amsmath,amsfonts,amssymb}
\usepackage{caption}
\usepackage{subcaption}
\usepackage{float}
\usepackage{multirow}
\usepackage{pifont}
\usepackage{threeparttable}
\usepackage{tablefootnote}

\usepackage{comment}
\usepackage[super]{nth}
\usepackage{balance}
\usepackage{xspace}

\newcommand{\ie}{i.e.,\xspace}

\newcommand{\eg}{e.g.,\xspace}

\newcommand{\etc}{etc.\xspace}
\newcommand{\wrt}{w.r.t.\xspace}
\newcommand{\system}{AGL\xspace}
\newcommand{\preprocessing}{GraphFlat\xspace}
\newcommand{\training}{GraphTrainer\xspace}
\newcommand{\infer}{GraphInfer\xspace}

\vldbTitle{AGL: a Scalable System for Industrial-purpose Graph Machine Learning}
\vldbAuthors{Dalong Zhang, Xin Huang, Ziqi Liu, Zhiyang Hu, Xianzheng Song, Zhibang Ge, Zhiqiang Zhang, Lin Wang, Jun Zhou, Yuan Qi}
\vldbDOI{https://doi.org/10.14778/xxxxxxx.xxxxxxx}
\vldbVolume{12}
\vldbNumber{xxx}
\vldbYear{2019}

\begin{document}

% ****************** TITLE ****************************************

\title{AGL: A Scalable System for Industrial-purpose Graph Machine Learning}

% possible, but not really needed or used for PVLDB:
%\subtitle{[Extended Abstract]
%\titlenote{A full version of this paper is available as\textit{Author's Guide to Preparing ACM SIG Proceedings Using \LaTeX$2_\epsilon$\ and BibTeX} at \texttt{www.acm.org/eaddress.htm}}}

% ****************** AUTHORS **************************************


\numberofauthors{1} %  in this sample file, there are a *total*

%\thanks{*Equal Contributions.}
%
%  \author{
%  \alignauthor
%  Dalong Zhang, Xin Huang, Ziqi Liu, Zhiyang Hu, Xianzheng Song\\
%  	\affaddr{Ant Financial Services Group, Hangzhou, China}\\
%	\email{\{dalong.zdl,huangxi.hx,ziqiliu,zhiyang.hzhy,xianzheng.sxz,\}@antfin.com}
% \alignauthor
%  }

\author{
% 1st. author
\alignauthor
Dalong Zhang, Xin Huang, Ziqi Liu, Zhiyang Hu, Xianzheng Song, Zhibang Ge, Zhiqiang Zhang, Lin Wang, Jun Zhou, Yang Shuang, Yuan Qi\\
       \affaddr{\ }\\
       \affaddr{Ant Financial Services Group, Hangzhou, China}\\
       \email{{dalong.zdl, huangxi.hx, ziqiliu, zhiyang.hzhy, xianzheng.sxz, zhibang.zg,\\lingyao.zzq, fred.wl, jun.zhoujun, shuang.yang, yuan.qi}@antfin.com}
% 2nd. author
%\alignauthor
%Xin Huang\\
%       \affaddr{Ant Financial Services Group}\\
%       \affaddr{Hangzhou, China}\\
%       \email{dalong.zdl@antfin.com}
%% 3rd. author
%\alignauthor  Ziqi Liu\\
%       \affaddr{Ant Financial Services Group}\\
%       \affaddr{Hangzhou, China}\\
%       \email{dalong.zdl@antfin.com}
%\alignauthor  Zhiyang Hu\\
%       \affaddr{Ant Financial Services Group}\\
%       \affaddr{Hangzhou, China}\\
%       \email{dalong.zdl@antfin.com}
%\and  
%\alignauthor  Xianzheng Song\\
%       \affaddr{Ant Financial Services Group}\\
%       \affaddr{Hangzhou, China}\\
%       \email{dalong.zdl@antfin.com}
%% use '\and' if you need 'another row' of author names
%% 4th. author
%\alignauthor  Zhibang Ge\\
%       \affaddr{Ant Financial Services Group}\\
%       \affaddr{Hangzhou, China}\\
%       \email{dalong.zdl@antfin.com}
%% 5th. author
%\alignauthor Zhiqiang Zhang\\
%       \affaddr{Ant Financial Services Group}\\
%       \affaddr{Hangzhou, China}\\
%       \email{dalong.zdl@antfin.com}
%% 6th. author
%\alignauthor Lin Wang\\
%       \affaddr{Ant Financial Services Group}\\
%       \affaddr{Hangzhou, China}\\
%       \email{dalong.zdl@antfin.com}
}
% There's nothing stopping you putting the seventh, eighth, etc.
% author on the opening page (as the 'third row') but we ask,
% for aesthetic reasons that you place these 'additional authors'
% in the \additional authors block, viz.

%\additionalauthors{Additional authors: Jun Zhou (Ant Financial Services Group, {\texttt{jsmith@affiliation.org}}), Yuan Qi (Ant Financial Services Group, {\texttt{jsmith@affiliation.org}})}

% Just remember to make sure that the TOTAL number of authors
% is the number that will appear on the first page PLUS the
% number that will appear in the \additionalauthors section.

\maketitle

\begin{abstract}
Machine learning over graphs have been emerging as powerful
learning tools for graph data. However, it is challenging
for industrial communities to leverage the techniques,
such as graph neural networks (GNNs), and solve real-world problems
at scale because of inherent data dependency in the graphs.
As such, we cannot simply train a GNN with classic
learning systems, for instance parameter server that assumes data 
parallel. Existing systems store the graph data in-memory
for fast accesses either in a single machine or 
graph stores from remote.
The major drawbacks are in three-fold. First, they 
cannot scale because of the limitations on the volume of the
memory, or the bandwidth between graph stores and workers.
Second, they require extra development of graph stores without well 
exploiting mature infrastructures such as MapReduce that guarantee
good system properties. Third, they focus
on training but ignore the optimization of inference over graphs, thus makes them 
an unintegrated system.

In this paper, we design \system, a scalable, fault-tolerance and integrated system,
with fully-functional training and inference for GNNs.
Our system design follows the message passing scheme underlying the
computations of GNNs. We design to generate the $k$-hop
neighborhood, an information-complete subgraph for each node, as well as do the
inference simply by merging values from in-edge neighbors and propagating
values to out-edge neighbors via MapReduce. In addition, the $k$-hop
neighborhood contains information-complete subgraphs for each node,
thus we simply do the training on parameter servers due to data 
independency. Our system \system, implemented
on mature infrastructures, can finish the training
of a 2-layer graph attention network on a graph with billions of nodes and hundred billions of edges in 14 hours,
and complete the inference in 1.2 hour.
\end{abstract}

\section{Introduction}
\label{sec:intro}

In recent years, both of the industrial and academic communities have paid much more attentions to machine learning over graph structure data.
The \emph{\textbf{G}raph \textbf{M}achine \textbf{L}earning} (abbreviated as GML) not only claims success in traditional graph mining tasks (e.g., node classifications~\cite{b1,b2,b3,b4}, link property
predictions~\cite{zhang2019dsslp} and graph property predictions~\cite{b6,b7}), but also brings great improvement to the tasks of other domains (\eg knowledge graph~\cite{b8,b9}, NLP~\cite{b10}, Computer Vision~\cite{b11,b12}, \etc). 
Besides, more and more Internet companies have applied the GML technique in solving various of industrial problems and achieved great success (\eg recommendation~\cite{b14,b14_2}, marketing~\cite{liu2019graph}, fraud detection~\cite{b15,hu2019cash}, loan default prediction~\cite{b16}, \etc).

In order to use graph machine learning
techniques to solve real-world problems by 
leveraging industrial-scale graphs, we are
required to build a learning system with \emph{scalibility},
\emph{fault tolerance}, and \emph{integrality} of fully-functional
training/inference workloads.
However, the computation graph of graph machine learning tasks
are fundamentally different from traditional learning
tasks because of \emph{data dependency}. That is, the 
computation graph of each sample is independent of
other samples in existing classic parameter server frameworks~\cite{b22}
assuming \emph{data parallel},
while the computation graph of each node in graph learning tasks
is dependent on the $k$-hop neighbors of that node.
The data dependency in graph learning tasks makes
that we can no longer store the training or inference samples
in disk and access them through pipelines~\cite{b22}. Instead, we 
have to store the graph data in-memory for fast data accesses.
This makes us fail to simply build a learning and inference system for graph
learning tasks based on existing parameter server frameworks 
that simply maintain the model consistency in parameter servers and 
do the workload with data parallel in each worker.

Several companies make efforts to design ingenious system architectures 
for various GML techniques. 
Facebook presents PyTorch-BigGraph (PBG)~\cite{b17}, a 
large-scale \emph{network embedding} system, which aims to 
produce unsupervised node embedding from multi-relation data. 
However, PBG is not suitable for plenty of real-world scenarios 
in which the graphs have rich attributes over nodes and 
edges (called \emph{attributed graph}). 
Recently, by leveraging deep learning techniques, 
\emph{graph neural network} (GNN) is able to generate high-quality 
representation for attributed graph, or build end-to-end 
supervised model over attributed graph with labeled nodes or 
edges, and thus becomes the more general-purpose GML technique.
For example, Deep Graph 
Library (DGL)~\cite{b18}, PyTorch Geometric (PyG)~\cite{b24} and AliGraph~\cite{b19} have been developed
for training graph neural networks over large-scale attributed 
graphs. Among of them, DGL and PyG are designed as a single-machine system
to deal with industrial-scale graphs in-memory based on a monster machine,
for instance, AWS x1.32xlarge with 2TB memory. As a distributed
system, AliGraph implements distributed in-memory graph storage engine,
which requires standalone deployment before training a GNN model.

However, the real industrial graph data could be huge. The social graph 
in Facebook\footnote{\url{https://en.wikipedia.org/wiki/Facebook,\_Inc.}} 
includes over two billion nodes and over a trillion edges~\cite{b17,b20}. 
The heterogenous financial graph in Ant 
Financial\footnote{\url{https://en.wikipedia.org/wiki/Alipay}} contains 
billions of nodes and  hundreds of billion edges with rich 
attribute information, as well as the e-commerce graph in 
Alibaba\footnote{\url{https://en.wikipedia.org/wiki/Alibaba\_Group}}.  
Table \ref{tab:summary_scale} summarizes the scale of graphs 
which is reported by several state-of-the-art GML systems.
The graph data at this scale may result into $100$ TB of data 
by counting features associated
with those nodes and edges. Those data are infeasible to be stored
in a single machine like DGL. Furthermore, the communications
between graph store engine storing the graphs and features
associated with nodes and edges, and workers could be very
huge. For instance, assuming a batch of a subgraph with $1000$ nodes
and $10000$ edges, this could result into $1$ MB of bulk to be communicated
between graph stores and workers, which is intolerant in case that we cannot
access the data in pipelines. Besides, this requires a well structured
network with high enough bandwidth. 

To summarize, firstly, existing industrial designs of learning systems
require the in-memory storage of graph data either in a single
monster machine that could not handle real industrial-scale graph data,
or in a customized graph store that could lead to
huge amount of communications between graph stores and workers.
This makes them \emph{not scale} to larger graph data. Second, they do
\emph{not well exploit the classic infrastructures}, such as MapReduce or
parameter servers, for fault tolerance purpose. Third, most
of existing frameworks pay more attention to training of graph learning
models, but \emph{ignore the system integrality}, for example,
the optimization of inference tasks while deploying graph machine learning models.

\begin{table}[t]
\caption{Graph scale reported by different GML systems.}
\begin{center}
\label{tab:summary_scale}
\begin{tabular}{c||cc}
\toprule
Indices         &\#Nodes      &\#Edges        \\
\midrule
DGL~\cite{b18}      &$5\times10^8$  &Unknown            \\
PBG~\cite{b17}      &$1.2\times10^8$     &$2.7\times10^9$              \\
Aligraph~\cite{b19}       &$4.9\times 10^8$    &$6.8\times 10^9$           \\
PinSage~\cite{b14_2}        &$3\times10^9$  &$1.8\times10^{10}$             \\
 \bottomrule
\end{tabular}
\end{center}
\end{table}

Take all those concerns into considerations, we build 
\system (\textbf{A}nt \textbf{G}raph machine \textbf{L}earning system), an integrated system for industrial-purpose 
graph learning. 
The key insight of our system design is based on the
\emph{message passing (merging and propagation)} scheme underlying the computation graph of
graph neural networks. 

In the phase of training graph neural networks, we 
propose to construct \emph{$k$-hop neighborhood} that provides information-complete
subgraphs for computing each node's $k$-hop embeddings based on
message passing by \emph{merging} neighbors from in-edges and 
\emph{propagating} merged information to neighbors along out-edges. 
The benefit of decomposing the original graph into tiny pieces of
subgraphs, i.e. $k$-hop neighborhood, is that the computation graph
of each node is \emph{independent} of other nodes \emph{again}. That means we can still
enjoy the properties of fault tolerance, flexible model consistency from classic
parameter server frameworks without extra effort on maintaining the graph 
stores~\cite{b19}. 

In the inference phase of graph neural networks,
we propose to split a well trained $K$-layer graph neural networks into $K$
slices plus one slice related to the prediction model. With the slices we do message passing by first merging the $k$-th 
layer embedding from each node's in-edge neighbors, then propagating embeddings
to their out-edge neighbors, with $k$ starts from $1$ to $K$.

We abstract all the message passing schemes in training and inference, and 
implement them simply using MapReduce~\cite{dean2008mapreduce}. 
Since both MapReduce and parameter servers have been developed as 
infrastructures commonly in industrial companies, our system for 
graph machine learning tasks can still benefit the properties like 
fault tolerance and scalibility even with commodity machines which 
is cheap and widely used. 
Moreover, compared with the inference
based on architectures like DGL and AliGraph, the 
implementation of our inference maximally utilizes each nodes'
embeddings, so as to significantly boost the inference jobs.
In addition, we propose several techniques to accelerate the floating
point calculations in training procedures from model level to operator level.
As a result, we successfully accelerate the training of GNNs in a single machine compared with DGL/PyG, and achieve a near-linear speedup with a CPU cluster in real product scenarios.

It's worthing noting that, when working on a graph with $6.23\times 10^9$ nodes and $3.38\times 10^{11}$ edges, \system can finish the training of a 2-layer GAT model with $1.2\times 10^8$ target nodes in 14 hours (7 epochs until convergence, 100 workers), and completes the inference on the whole graph in only 1.2 hours. 
To our best knowledge, this is the largest-ever application of graph embeddings and proves the high scalability and efficiency of our system in real industrial scenarios.

\section{Preliminaries}
\label{sec:preliminaries}
In this section, we introduce some notations, and highlight the fundamental
computation paradigm, i.e. message passing, in graph neural networks (GNN). Finally, we introduce the concept of $K$-hop neighborhood 
to help realize the data independency in graph learning tasks. Both of the abstraction of message passing scheme and 
$K$-hop neighborhood play an important role in the design of our system.

\subsection{Notations}\label{sec:notation}
A \emph{directed and weighted attributed graph} can be defined as
$\mathcal{G}=\{\mathcal{V},\mathcal{E},\mathbf{A},\mathbf{X},\mathbf{E}\}$, where
$\mathcal{V}$ and $\mathcal{E} \in \mathcal{V} \times \mathcal{V}$ are the
node set and edge set of $\mathcal{G}$, respectively.
$\mathbf{A} \in \mathbb{R}^{|\mathcal{V}|\times|\mathcal{V}|}$ is the sparse weighted adjacent matrix such that its element $\mathbf{A}_{v,u}>0$ represents the weight of a directed edge from node $u$ to node $v$ (\ie $(v,u) \in \mathcal{E}$), and $\mathbf{A}_{v,u}=0$ represents there is no edge (\ie $(v,u) \notin \mathcal{E}$).
$\mathbf{X} \in \mathbb{R}^{|\mathcal{V}| \times f^n}$ is a matrix consisting of all nodes' $f^n$-dimensional feature vectors,
and $\mathbf{E} \in \mathbb{R}^{|\mathcal{V}| \times |\mathcal{V}| \times f^e}$ is a sparse tensor consisting of all edges' $f^e$-dimensional feature vectors.
Specifically, $\mathbf{x}_v$ denotes the feature vector of $v$, $\mathbf{e}_{v,u}$ denotes the feature vector of edge $(v,u)$ if $(v,u) \in \mathcal{E}$, otherwise $\mathbf{e}_{v,u}=\mathbf{0}$.
In our setting, an \emph{undirected graph} is treated as a special directed
graph, in which each undirected edge $(v,u)$ is decomposed as two directed
edges with the same edge feature, \ie $(v,u)$ and $(u,v)$.
Moreover, we use $\mathcal{N}^+_v$ to denote the set of nodes
directly pointing at $v$, \ie $\mathcal{N}^+_v = \{u : \mathbf{A}_{v,u}>0\}$,
$\mathcal{N}^-_v$ to denote the set of nodes directly pointed
by $v$, \ie $\mathcal{N}^-_v = \{u : \mathbf{A}_{u,v}>0\}$,
and $\mathcal{N}_v = \mathcal{N}^+_v \cup \mathcal{N}^-_v$. In other words,
$\mathcal{N}^+_v$ denotes the set of in-edge neighbors of $v$, while 
$\mathcal{N}^-_v$ denotes the set of out-edge neighbors of $v$.
We call the edges pointing at a certain node as its \emph{in-edges}, while the
edges pointed by this node as its \emph{out-edges}.

\subsection{Graph Neural Networks}\label{sec:gml}
Most GML models aim to encode a graph structure 
(\eg node, edge, subgraph or the entire graph) 
as a low dimensional embedding, which is used as the 
input of the downstream machine learning tasks, in an 
end-to-end or decoupled manner.  The proposed \system mainly 
focuses on GNNs, which is a category of GML models widely-used.
Each layer of GNNs generates the intermediate embedding by 
aggregating the information of target node's in-edge neighbors. 
After stacking several GNN layers, we obtain the final 
embedding, which integrate the entire receptive field of the targeted node. 
Specifically, we denote the computation paradigm of the
$k$\textsuperscript{th} GNN layer as follows:
\begin{equation} \label{eq:gnn}
\mathbf{h}_v^{(k+1)} = \phi^{(k)}(\{\mathbf{h}_i^{(k)}\}_{i \in \{v\} \cup \mathcal{N}^+_v}, \{\mathbf{e}_{v,u}\}_{\mathbf{A}_{v,u}>0}; \mathbf{W}_{\phi}^{(k)}), 
\end{equation}
where $\mathbf{h}_v^{(k)}$ denotes node $v$'s intermediate 
embedding in the $k$\textsuperscript{th} layer and $\mathbf{h}_v^{(0)} = \mathbf{x}_v$. 
The function $\phi^{(k)}$ parameterized by $\mathbf{W}_{\phi}^{(k)}$, takes the 
embeddings of $v$ and its in-edge neighbors $\mathcal{N}^+_v$, as well as 
the edge features associated with $v$'s in-edges as inputs, and outputs 
the embedding for the next GNN layer.

The above computations of GNNs can be formulated
in the message passing paradigm.
That is, we collect keys (i.e., node ids)
and their values (i.e., embeddings). We first merge all
the values from each node's in-edge neighbors so as to have
the new values for the nodes. After that, 
we propagate the new values to destination nodes
via out-edges.
After $K$ times of such merging and propagation,
we complete the computation of GNNs.
We will discuss in the following sections that
such a paradigm will be generalized to the
training and inference of GNNs.

\subsection{K-hop Neighborhood}\label{sec:khop}
\begin{definition}\label{def:khop}
\emph{\textbf{k-hop neighborhood}}.
The $k$-hop neighborhood \wrt a targeted node $v$, denoted as 
$\mathcal{G}_v^k$, is defined as the \emph{induced attributed subgraph} 
of $\mathcal{G}$ whose node set is
$\mathcal{V}_v^k = \{v\} \cup \{u: d(v,u) \leq k\}$, where $d(v,u)$ denotes 
the length of the shortest path from $u$ to $v$. 
Its edge set consists of the edges in $\mathcal{E}$ that 
have both endpoints in its node set, i.e. $\mathcal{E}_v^k = \{(u,u'): (u,u')\in \mathcal{E} 
\land u\in\mathcal{V}_v^k \land u'\in\mathcal{V}_v^k \}$. 
Moreover, it contains the feature vectors of the nodes and edges 
in the $k$-hop neighborhood, $\mathbf{X}_v^k$ and $\mathbf{E}_v^k$. 
Without loss of generality, we define the $0$-hop neighborhood 
\wrt $v$ as the node $v$ itself.
%consists of the targeted 
%node $v$ and other nodes at a distance of not larger than $k$.
%Specifically, the node set of the $k$-hop neighborhood \wrt $v$ is 
%denoted as $\{v\} \cup \{u: d(v,u) \leq k\}$, where $d(v,u)$ denotes 
%the number of edges in the shortest path between $v$ and $u$. 
\end{definition}

The following theorem shows the connection between the  
computation of GNNs and the k-hop neighborhood.

\begin{theorem}\label{the:khop}
Let $\mathcal{G}_v^k$ be the k-hop neighborhood \wrt the target node $v$, then $\mathcal{G}_v^k$ contains the \textbf{sufficient and necessary information} for a $k$-layer GNN model, which follows the paradigm of \autoref{eq:gnn}, to generate the embedding of node $v$.
\end{theorem}

First, the $0$\textsuperscript{th} layer embedding is directly assigned by the raw feature, \ie $\mathbf{h}_v^{(0)} = \mathbf{x}_v$, which is also the $0$-hop neighborhood.
And then, from \autoref{eq:gnn}, it's easy to find that the output embedding of $v$ in each subsequent layer is generated only based on the embedding of 
the 1-hop in-edge neighbors \wrt $v$ from the previous layer.
Therefore, by applying mathematical induction, it's easy 
to prove the \autoref{the:khop}.  Moreover, we can extend 
the theorem to a batch of nodes, that is the intersection of 
the $k$-hop neighborhoods \wrt a batch of nodes provides 
the sufficient and necessary information for a $k$-layer GNN  
model to generate the embedding of all nodes in the batch. 
This simple theorem implies that in a $k$-layer GNN model 
the target node's embedding at the $k^{\text{th}}$ layer 
only depends on its $k$-hop neighborhood, rather than the entire graph.

\section{System} \label{sec:system}
In this section, we first give an overview of our \system system.
Then, we elaborate three core modules, \ie \preprocessing,  
\training and \infer.
At last, we give a demo example on how to implement a simple GCN model with the proposed \system system.%discuss the system-provided APIs for the development of GNN models. 

\subsection{System Overview}
Our major motivation of building \system is that the industrial communities desiderate
an \emph{integrated} system of fully-functional training/inference over graph data,
with \emph{scalability}, and in the meanwhile has the properties of \emph{fault tolerance} based
on \emph{mature industrial infrastructures} like MapReduce, parameter servers, etc.
That is, instead of requiring a single monster machine or customized graph stores
with huge memory and high bandwidth networks, which could be expensive for Internet
companies to upgrade their infrastructures, we sought to give a solution
based on mature and classic infrastructures, which is ease-to-deploy
while enjoying various properties like fault tolerance and so on. Second, we need
the solution based on mature infrastructures scale to industrial-scale graph data. Third, besides the optimization of training, we aim to boost the inference
tasks over graphs because labeled data are very limited (say ten million) in practice compared
with unlabeled data, typically billions of nodes, to be inferred.

The principle of designing \system is based on the message passing scheme
underlying the computations of GNNs.
That is, we first merge all the informations from each node's in-edge
neighbors, and then propagate those merged informations to
the destination nodes via out-edges.
We repeatedly apply such a principle to the training and inference processes,
and develop \emph{GraphFlat} and \emph{GraphInfer}. Basically, GraphFlat is to generate
independent $K$-hop neighborhoods in the training process, while GraphInfer
is to infer nodes' embeddings given a well trained GNN model.

Based on the motivation and design principle, the proposed \system leverages several powerful parallel architectures, such as MapReduce and Parameter Server, to build each of its components with exquisitely-designed distributed implementations.  
As a result, even being deployed on the clusters with machines that has relatively low computing capacity and limited memory, \system gains comparable effectiveness and higher efficiency against several state-of-the-art systems.
Moreover, it has the ability to perform fully-functional graph machine learning over industrial-scale graph with billions of nodes and hundred billions of edges.

\begin{figure}[htbp]
\centering
\includegraphics[width=0.98 \linewidth]{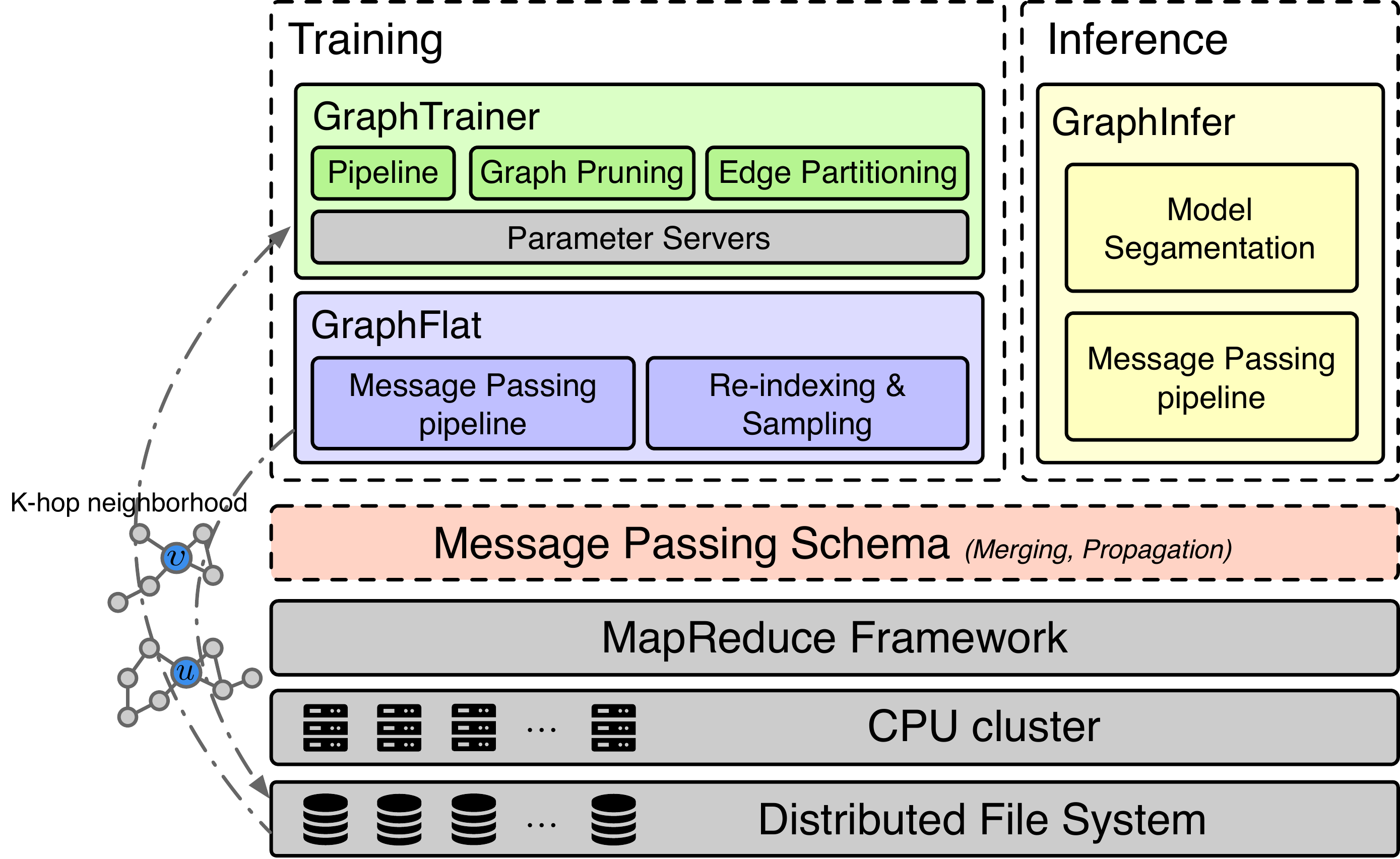}%{Fg1_system_overview_0304.eps}
\caption{System architecture of \system. }
\label{fig:ALPS-GraphML_framework}
\end{figure}

Figure~\ref{fig:ALPS-GraphML_framework} depicts the system architecture of \system, which consists of three modules:
 
 (1) \textbf{\preprocessing}.
\preprocessing is an efficient and distributed generator, based on message passing,
for generating $K$-hop neighborhoods that contain information complete subgraphs
of each targeted nodes.
Those tiny $k$-hop neighborhoods are flattened to a protobuf
strings\footnote{\url{https://en.wikipedia.org/wiki/Protocol_Buffers}}
and stored on a distributed file system.
Since the $k$-hop neighborhood contains sufficient and necessary
information for each targeted node, we can load one or a batch of
them rather than the entire graph into memory, and do the training
similar to any other traditional learning methods.
Besides, we propose a re-indexing technique together with a
sampling framework to handle ``hub'' nodes in real-world applications.
Our design is based on the observation that the amount of labeled
nodes is limited, and we can store those $K$-hop neighborhoods
associated with the labeled nodes in disk without too much cost.

 (2) \textbf{\training}.
Based on the data independency guaranteed by \preprocessing,
\training leverages many techniques, such as pipeline, pruning, and
edge-partition, to eliminate the overhead on I/O and optimize the floating point
calculations during the training of GNN models.
As a result,  \training gains a high near-linear speedup in real
industrial scenarios even on a generic CPU cluster with commodity machines.

(3) \textbf{\infer}.
We develop GraphInfer, a distributed inference module that splits $K$ layer GNN models
into $K$ slices, and applies the message passing $K$ times based on MapReduce.
GraphInfer maximally utilizes the embedding of each node because all the intermediate
embedding at the $k$-th layer will be propagated to next round of message passing.
This significantly boosts the inference tasks.

%(2) \emph{Training module}. 
%The training module is to distributedly train a certain graph machine %learning model (like node classification model, link prediction model). 
%Based on GraphFlat, the training phase is parallelly performed on %multi-machines in data-parallel pattern based on the Parameter Server architecture.
%As the efficiency matters in some product scenarios, especially when applications in those scenarios have high request on timeliness, we leverages many techniques to optimize
%the training procedure in different levels from model level to operator level.

%Prepare data from the distributed file system and train a general graph embedding model;

%(3) \emph{Inference module}. 
%The inference module is to deploy a well-trained model and perform the inference procedure.
%However, as the test set is usually much larger than the training set, the inference could be very slow, let alone in real industrial scenarios.
%Also, we develop a model-split technique and build a MapReduce pipeline to perform the inference procedure on huge graphs, and achieve a significant improvement compared with the original one.
%Deploy a well-trained model and perform the inference.

Details about our system will be presented in the following sections.
%These three modules are carefully designed, which makes our system gain the gift to scale to huge graphs and run parallelly. 
%In the data preprocessing module, huge graphs are partitioned into small sets of $k$-hop neighborhoods, which can be loaded into commodity machines and contain complete information for computing $k$-hop embeddings.

%Moreover, the data preprocessing and inference module are designed based on MapReduce, while the model training module is deployed on a distributed cluster base on the parameter-server architecture, which means that every module hold the ability to operate in a parallel mode and makes our system efficient to process graphs with huge scale. Our system will be detailed in the following.

\subsection{\preprocessing: Distributed Generator of $k$-hop Neighborhood}
\label{sec:graphflat}

The major issue of training graph neural networks is 
the inherent data dependency among graph data. To do
the feedforward computation of each node, we have to
read its associated neighbors and neighbors' neighbors,
and so on so forth. This makes us fail to deploy such
network architecture simply based on existing parameter server
learning frameworks that assumes data parallel. Moreover, 
developing extra graph stores for query of each node's 
subgraphs is expensive to most of industrial companies.
That is, such a design would not benefit us with existing 
commonly deployed infrastructures that are mature and 
have various properties like fault tolerance.

%One of the most challenging issues of industrial GML system is how to scale to industrial graphs with billions of nodes and hundreds of billion edges.
%The memory cost to load such a graph may be over 10 terabytes.
%Existing solutions may employ a monster machine or even build a cluster with massive of memory to handle the huge graph.
%However, as there may exist hundreds of GML applications in an Internet company every day, those solutions are too expensive for the real-world scenario.
%Thus, we are seeking a cost-effective architecture, to empower the clusters of commodity machines, which exists in most Internet company, with the ability of performing GML applications. 

Fortunately, according to \autoref{the:khop}, the $k$-hop neighborhood \wrt a
target node provides sufficient and necessary information to generate
the $k$\textsuperscript{th}-layer node embedding.
Therefore, we can divide a industrial-scale graph into massive of
tiny $k$-hop neighborhoods \emph{w.r.t.} their target nodes in
advance, and load one or a batch of them rather than the entire
graph into memory in the training phase.
Following this idea, we develop \preprocessing, an efficient distributed
generator of $k$-hop neighborhood. Moreover, we further introduce a 
re-indexing strategy and design a sampling framework to handle ``hub'' nodes and ensures 
the load balance of \preprocessing.
The details are presented as follows.

\begin{figure*}[htbp]
\centering
\includegraphics[width=0.98 \linewidth]{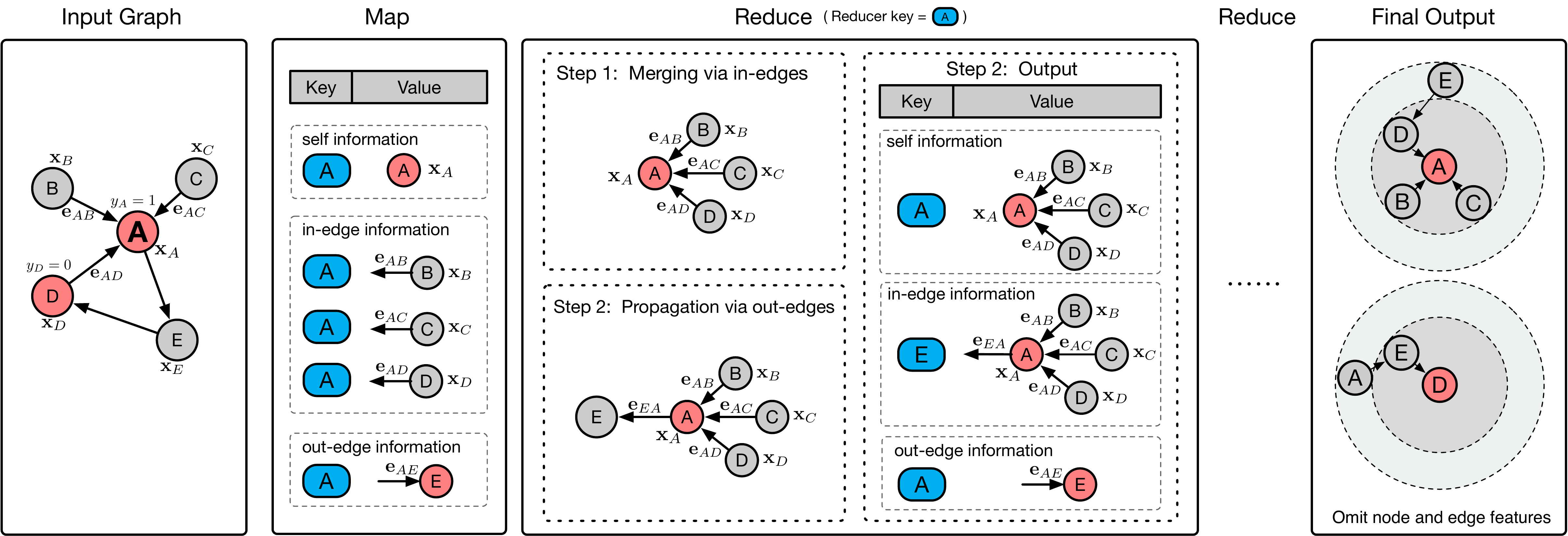}%{Fg2_graphflat_0304.eps}
\caption{The pipeline of \preprocessing. }
\label{fig:graphflat}
\end{figure*}

\subsubsection{Distributed pipeline to generate $k$-hop neighborhood} 

In this section, we design a distributed pipeline to generate 
$k$-hop neighborhoods in the spirit of message passing, 
and implement it with MapReduce infrastructure.

%A major problem in GraphFlat module is how to extract $k$-hop neighborhood from large scale graphs in industrial scenarios. 
%It's easy to perform the extraction phase on small graphs with commodity machines in standalone mode.
%However, when scale to huge graphs with billions of nodes and hundreds of billion edges, such extracting procedure will meet its bottleneck in both efficiency and memory-cost.
%Hence, we develop a MapReduce pipeline to distributedly perform the extracting procedure.

\autoref{fig:graphflat} illustrates the workflow of the proposed pipeline.
The key insight behind is that, for a certain node $v$, we first 
\emph{receive and merge} the information from the 
in-edge neighbors $\mathcal{N}_v^{+}$ pointing at $v$, 
then \emph{propagate} the merged results to the out-edge neighbors
$\mathcal{N}_v^{-}$ pointed by $v$. 
By repeating this procedure $k$ times, we finally get 
the $k$-hop neighborhoods.

Assume that we take a \emph{node table} and an \emph{edge table} as input.
Specifically, the \emph{node table} consists of node ids and node features,
while the \emph{edge table} consists of source node ids, destination node ids and the edge features.
The overall pipeline to generate the $K$-hop neighborhood can be summarized as follows:
\begin{enumerate}
\renewcommand{\labelenumi}{(\theenumi)}
\item \textbf{Map}.
The \emph{Map} phase runs only once at the beginning of the pipeline.
For a certain node, the \emph{Map} phase generates three kinds of information, \ie the self information (\ie node feature), the in-edge information (\ie feature of the in-edge and the neighbor node) and the out-edge information (\ie feature of the out-edge).
Note that we set the node id as the \emph{shuffle key} and the various information as the \emph{value} for the following \emph{Reduce} phase.

\item \textbf{Reduce}.
The \emph{Reduce} phase runs $K$ times to generate the $K$-hop neighborhood.
In the $k$\textsuperscript{th} round, a reducer first collects all \emph{values} (\ie three kinds of information) with the same \emph{shuffle key} (\ie the same node ids),
then \emph{merges} the self information and the in-edge information as its new self information.
Note that the new self information become the node's $k$-hop neighborhood.
Next, the new self information is \emph{propagated} to other destination nodes pointed along the out-edges, and is used to construct the new in-edge information \wrt the destination nodes.
All of the out-edge information remain unchanged for the next \emph{reduce} phase.
At last, the reducer outputs the new data records, with the node ids and the updated information as the new \emph{shuffle key} and \emph{value} respectively, to the disk.

\item \textbf{Storing}.
After $k$ \emph{Reduce} phase, the final self information becomes the $k$-hop neighborhood.
We transform the self information of all targeted nodes into the protobuf strings and store them into the distributed filesystem.
\end{enumerate}

Throughout the MapReduce pipeline, the key operations 
are \emph{merging} and \emph{propagation}.
In each round, given a node $v$, we merge its 
self information and in-edge information from last round, 
and the merged results serve as the self information of $v$.
We then propagate the new self information via out-edges to the destination nodes.
At the end of this pipeline, the $k$-hop neighborhood \wrt a certain targeted node is flattened to a protobuf string.
That's why we call this pipeline \textbf{\preprocessing}.
%\preprocessing finally transform data records into protobuf strings and store them into a distributed file system for downstream training phase.
%%%%%%%%commented by zdl 0302
% A toy example of \preprocessing's final outputs for a node classification task is shown in \autoref{fig:graphflat_output}.
Note that, since the $k$-hop neighborhood \emph{w.r.t.} to a node helps discriminate the node from others, we also call it \emph{GraphFeature}.

\subsubsection{Sampling \& Indexing}
\label{sec:sampling}
The distributed pipeline described in the previous subsection works well in most cases.
However, the degree distribution of the graphs can be skewed due to the existence
of ``hub'' nodes, especially in the industrial scenario. This makes some of $k$-hop neighborhoods may cover
almost the entire graph. On one hand, in the \emph{Reduce} phase
of \preprocessing, reducers that process such ``hub'' nodes could be
much slower than others thus damage the load balances of \preprocessing.
On the other hand, the huge $k$-hop neighborhoods \wrt those ``hub'' nodes
may cause the Out Of Memory (OOM) problem in both \preprocessing and
the downstream model training. Moreover, the skewed data may also
lead to a poor accuracy of the trained GNN model.
Hence, we employ the re-indexing strategy and design a sampling framework for reducer in \preprocessing.
%% some problem in real industrial scenarios:
%(1) In reduce phase, the reducers that process those super-hub nodes will be far slower than others and block the full MapReduce pipeline.
%%OOM (Out Of Memory) problem in both extracting phase and the downstream training phase. 
%(2) The huge $k$-hop neighborhoods of those super-hub nodes may cause OOM (Out Of Memory) problem in both extracting phase and the downstream training phase.
%Moreover, in some applications, especially with high request on timelines, a node with its full $k$-hop neighborhood may lead to a high latency time, especially when there are some super-hub nodes.
%%In some applications, 
%%especially with high request on timelines, a node with its full $k$-hop neighborhood may lead to a high latency time, especially when there are some super-hub nodes.
%Hence, we leverage several techniques in designing a sampling strategy to solve those those problem:

\begin{figure}[h]
\centering
\includegraphics[width=0.9\linewidth]{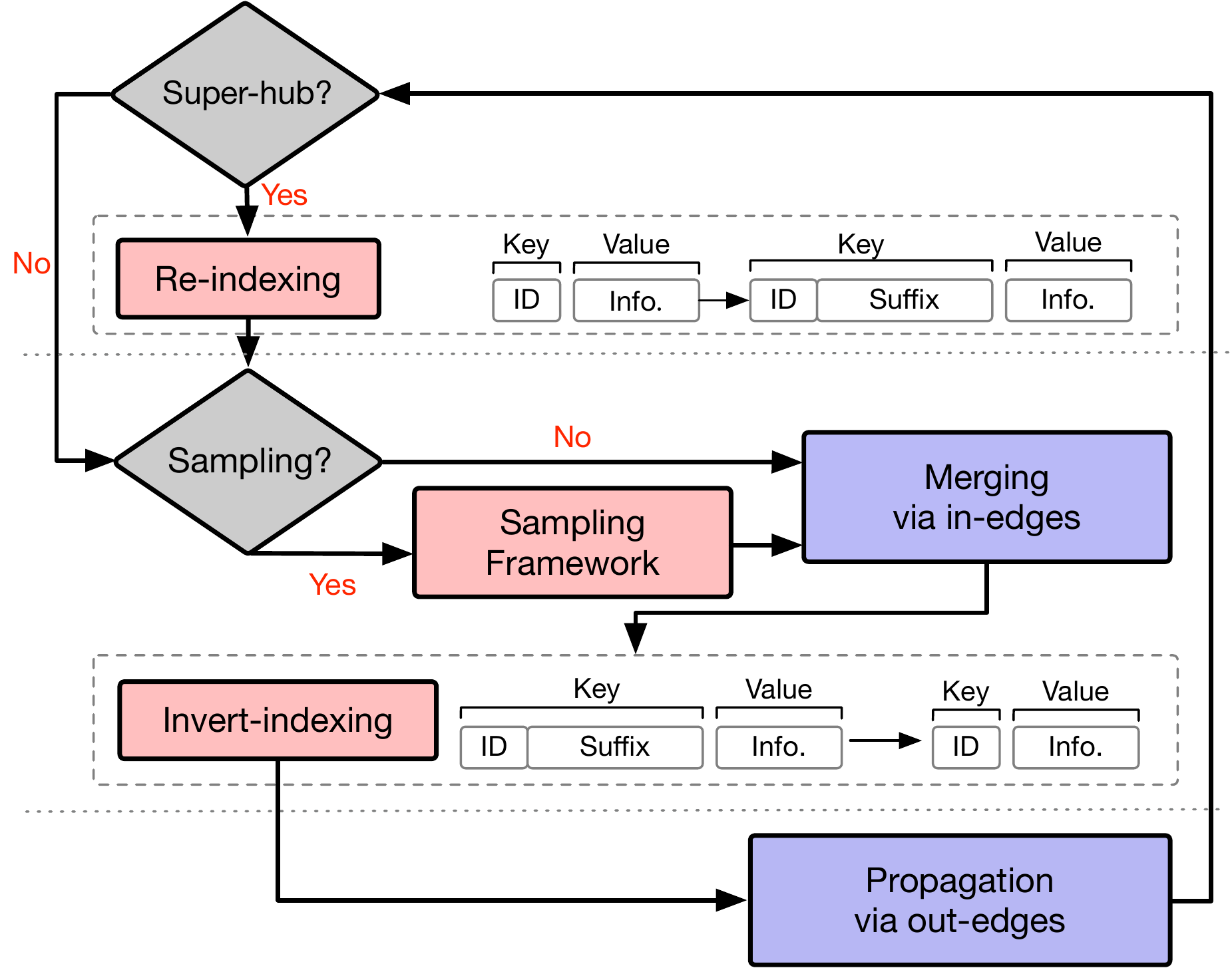}%{Figure4_sampling_indexing_0304.eps}
\caption{Workflow of sampling and indexing in \preprocessing.}
\label{fig:sampling_reducer}
\end{figure}

\begin{figure*}[t]
\centering
\includegraphics[width=0.98\linewidth]{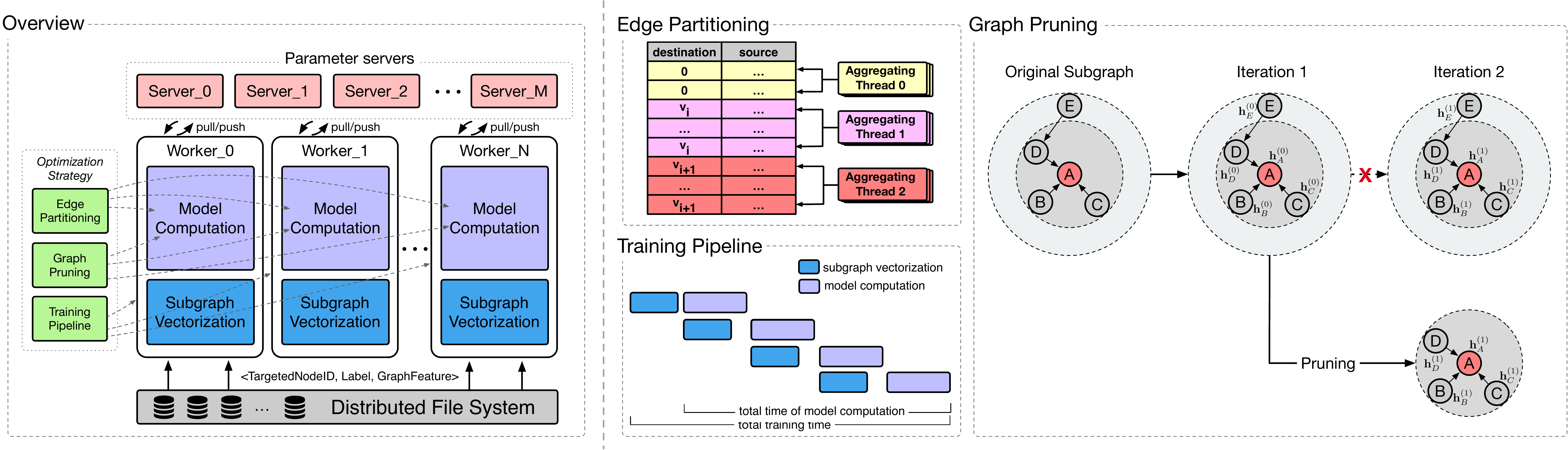}%{training_fig4_7_0304.eps}
\caption{Training workflow and optimization strategies. }
\label{fig:train_work_flow}
\end{figure*}

\autoref{fig:sampling_reducer} illustrates the
reducer with re-indexing and sampling strategies in \preprocessing.
Three key components of performing re-indexing and sampling are introduced as follows:
\begin{itemize}
\item \textbf{Re-indexing}.
When the in-degree of a certain \emph{shuffle key}
(\ie node id) exceeds a pre-defined threshold (like 10k),
we will update \emph{shuffle keys} by appending random
suffixes, which is used to randomly partition the data records
with the original \emph{shuffle key} into smaller pieces.
\item \textbf{Sampling framework}.
We build a distributed sampling framework and implement a set of sampling strategies (\eg uniform sampling, weighed sampling), to reduce the scale of the $k$-hop neighborhoods, especially for those ``hub'' nodes.
\item \textbf{Inverted indexing}. This component is responsible for replacing the reindexed \emph{shuffle key} with the original \emph{shuffle key}.
After that, the data records are outputted to the disk waiting for the downstream task.
\end{itemize}

Before sampling, the \emph{re-indexing} component is to
uniformly map data records associated with the same ``hub'' node
to a set of reducers.
It helps alleviate the load balance problem that could be caused by those ``hub'' nodes.
Then the sampling framework samples
a potion of the data records \emph{w.r.t.} a \emph{shuffle key}.
After that, the merging and propagation operation are performed as the original \emph{Reducer} does.
%Since different GML applications may have different demands on the sampling strategies, we leverage several techniques, such as rejection sampling, to meet those demands efficiently. 
%We also implement some priority-based method, such as sampling according to the weight of edges or nodes.
Next, the \emph{inverted indexing} component will recover the reindexed \emph{shuffle key} as the original \emph{shuffle key} (\ie node id) for the downstream task.
%After that, the downstream `Reduce' phase will load data according with the those recovered keys and broadcast the merged information through their out-edges.
%At the end of this `Reduce' phase we will perform the first stage to process super-hub nodes and output data records for the next cycle.

With re-indexing we make the process of ``hub'' nodes being partitioned
over a set of reducers, thus well maintain the load balances. With
sampling, we make the scale of $k$-hop neighborhoods being decreased
to an acceptable size.

\subsection{\training: Distributed Graph Training Framework} \label{sec:training}
In order to perform efficient training of $k$-hop neighborhoods generated by \preprocessing, we implement \training, the distributed graph training framework shown in \autoref{fig:train_work_flow}. 
The overall architecture of \training follows the parameter server design, which consists of two sets of components: the workers that perform the bulk of computation during model training, and the servers that maintain the current version of the graph model parameters.
Since the $k$-hop neighborhood contains sufficient and necessary information to train the GNN model,  the training workers of \training become independent of each other. 
They just have to process their own partitions of training data, and do not need extra communication with other workers.
Therefore, the training of a GNN model becomes similar to the training of a conventional machine learning model, in which the training data on each worker is self-contained.
Moreover, since most $k$-hop neighborhoods are tiny subgraph taking little memory footprint, training workers in \training only require to be deployed  on the commodity machines with limited computation resources (\ie CPU, memory, network bandwidth).

Considering the property of $k$-hop neighborhood as well as the characteristics of GNN training computation, we propose several optimization strategies, including training pipeline, graph pruning and edge partitioning, to improve the training efficiency. 
The rest of this subsection first introduce the overall training workflow, and then elaborate several graph-specific optimization strategies.

\subsubsection{Training workflow}
As shown in \autoref{fig:train_work_flow}, the training workflow mainly includes two phases, \ie subgraph vectorization and model computation.
We take the node classification task as an example to illustrate the two phases. 
In the node classification task, a batch of training examples can be formulated as a set of triples $\mathcal{B}=\{<TargetedNodeId, Label, GraphFeature>\}$.
Different from the training process of the conventional machine learning models, which directly performs model computation, the training process of GNNs has to merge the subgraphs described by \emph{GraphFeatures} together, and then vectorize the merged subgraph as the following three matrices.
\begin{itemize}
\item \textbf{Adjacency matrix}: $\mathbf{A}_{\mathcal{B}}$.  A sparse matrix with nodes and edges of the merged subgraph. Edges in the sparse matrix are sorted by their destination nodes.
\item \textbf{Node feature matrix}: $\mathbf{X}_{\mathcal{B}}$. A matrix to record the features of all nodes in the merged subgraph.
\item \textbf{Edge feature matrix}: $\mathbf{E}_{\mathcal{B}}$. A matrix to record the features of all edges in the merged subgraph.
\end{itemize}
%\textbf{Node-level batching strategy.} Different from other GML systems (like PyG and DGL), which batch small graphs during training phase (that is, if a node is in a certain batch, all nodes and edges from the graph containing this node must be in this batch too), our system supports node-level batches based, no matter whether those nodes are in a same graph or not.
%For example, given a arbitrary set of ids together with their \emph{GraphFeature}s as a mini-batch, our system first merges those \emph{GraphFeature}s and then transforms the merged \emph{GraphFeature} into three matrixes: 
%\begin{itemize}
%\item \textbf{Adjacency matrix.}  A sparse matrix to record edges in subgraphs extracted from root nodes. Edges in the sparse matrix are sorted by their destination nodes.
%\item \textbf{Node feature.} A matrix to record node features.
%\item \textbf{Edge feature.}  (Optional) A matrix to record edge features.
%\end{itemize}
Note that these three matrices contain all information of the $k$-hop neighborhood \wrt all targeted nodes in $\mathcal{B}$.
They will be fed to the model computation phase, together with the node ids and labels. 
Based on the three matrices as well as the ids and labels of targeted nodes, the model computation phase is responsible for performing the forward and backward calculations. 

\subsubsection{Optimization strategies}
In this subsection, we will elaborate three graph-specific optimization strategies in different level, to boost the training efficiency. 
That is training pipeline (batch-level), graph pruning (graph-level) and edge partitioning (edge-level).

%\begin{figure}[htbp]
%\centering
%\includegraphics[width=0.80\linewidth]{optimization_level_0211}
%\caption{Three level optimization. }
%\label{fig:opt_level}
%\end{figure}

%\textbf{Pipelined training procedure.}
 \textbf{Training pipeline}. 
During GNN model training, each worker first read a batch of its training data from the disks, then it perform \emph{subgraph vectorization} and \emph{model computation}.
Performing these steps sequentially is time-consuming. 
Addressing this problem, we build a pipeline that consists of two stages: preprocessing stage including data reading and subgraph vectorization, and model computation stage. 
%Note that for most GNN models, the two stages take almost the same processing time. 
%Thus, we 
%The batching stage prepares batched data for the second stage, while the algorithm-performing stage run a forward propagation to generate embedding together with a backward propagation to compute gradients to update parameters. 
The two stages operate in a parallel manner.
Since the time consumed by preprocessing stage is relatively shorter that of the model computation stage, after several rounds, the total training time is nearly equal to that of performing model computation only.
 %For a certain model, the train module first read and preprocess a batch of data from distributed system and then perform a certain algorithm on those
 %For a single worker, during training procedure, data are loaded from a distributed file system and passed from several preprocess steps (i.e. user-defined parsing method, sampling method) before being fed to the model, which may block the training procedure and can not make full use of CPUs. 
 \begin{comment}
\begin{figure}[htbp]
\centering
\includegraphics[width=0.80 \linewidth]{fig5_training_pipeline.eps}
\caption{Two stage training pipeline. }
\label{fig:pipeline}
\end{figure}
\end{comment}
%To solve this problem, we build a pipelined training procedure as shown in Fig.~\ref{fig:pipeline}. The first several stages prepare data for the model-computing stage, while the model-computing stage run a forward propagation to generate embedding together with a backward propagation to compute gradients to update parameters. All those stages operate in a parallel manner and after several steps, the total time of training procedure is nearly equal to that of model-computing stage in the best situation.

%\textbf{Pruning strategy.} 
\textbf{Graph pruning}. 
%As mentioned above, the three matrices, \ie $\mathbf{A}_{\mathcal{B}}$, $\mathbf{X}_{\mathcal{B}}$ and $\mathbf{E}_{\mathcal{B}}$, contain sufficient and necessary information to generate node embedding for all targeted nodes in .
Given the three matrices $\mathbf{A}_{\mathcal{B}}$, $\mathbf{X}_{\mathcal{B}}$ and $\mathbf{E}_{\mathcal{B}}$ \wrt batch $\mathcal{B}$, we revise \autoref{eq:gnn} \wrt $\mathcal{B}$ as follows:
\begin{equation}\label{eq:gnn_batch}
\mathbf{H}_{\mathcal{B}}^{(k+1)} = \Phi^{(k)}(\mathbf{H}_{\mathcal{B}}^{(k)}, \mathbf{A}_{\mathcal{B}}, \mathbf{E}_{\mathcal{B}}; \mathbf{W}_{\Phi}^{(k)}), 
\end{equation}
where $\mathbf{H}_{\mathcal{B}}^{(k)}$ denotes the $k$\textsuperscript{th}-layer  intermediate embeddings of all nodes that appear in the $k$-hop neighborhood \wrt all targeted nodes in $\mathcal{B}$, and $\Phi^{(k)}$ denotes the aggregating function of the $k$\textsuperscript{th} layer.
We assume that the final embedding is the $K$\textsuperscript{th}-layer embedding, \ie $\mathbf{H}_{\mathcal{B}}^{(K)}$.
%Taking GCN \cite{} as an example, the \autoref{eq:gnn_batch} \wrt GCN is implemented as follows:
%\begin{equation}\label{eq:gcn_batch}
%\mathbf{H}_{\mathcal{B}}^{(k+1)} = \sigma(\tilde{\mathbf{A}}_{\mathcal{B}}\mathbf{H}_{\mathcal{B}}^{(k)}\mathbf{W}_{gcn}^{(k)}),
%\end{equation}
%where $\tilde{\mathbf{A}}_{\mathcal{B}}=\hat{\mathbf{D}}_{\mathcal{B}}^{-\frac{1}{2}} \hat{\mathbf{A}}_{\mathcal{B}} \hat{\mathbf{D}}_{\mathcal{B}}^{-\frac{1}{2}}$, $\hat{\mathbf{A}}_{\mathcal{B}} = \mathbf{A}_{\mathcal{B}}+\mathbf{I}$, $\hat{\mathbf{D}}_{\mathcal{B}}$ is the diagonal node degree matrix of $\hat{\mathbf{A}}_{\mathcal{B}}$, and $\sigma$ indicates a non-linear activation function.

However, \autoref{eq:gnn_batch} contains many unnecessary computations. 
On one hand, only the targeted nodes of $\mathcal{B}$ are labeled. Their embedding will be fed to the following part of the model. That means other embeddings in $\mathbf{H}_{\mathcal{B}}^{(K)}$ are unnecessary to the following part of the model. 
On the other hand, the three matrices $\mathbf{A}_{\mathcal{B}}$, $\mathbf{X}_{\mathcal{B}}$ and $\mathbf{E}_{\mathcal{B}}$  can provide sufficient and necessary information only for the targeted nodes.
Thus other embeddings in $\mathbf{H}_{\mathcal{B}}^{(K)}$ could be generated incorrectly due to the lack of sufficient information. 

%As stated before, a certain $k$-hop neighborhood only ensures that the root node hold an information-sufficient surrounding environment the same as that from the full graph, when computing $k$-hop embeddings. And in GraphFlat module, only the root node is labeled. Therefore, there's many unnecessary computations if we still perform original GML algorithm on all nodes of the $k$-hop neighborhood. 
Tackling this problem, we propose a \emph{graph pruning} strategy to reduce the unnecessary computations mentioned above. 
Given a targeted node $v$, for any node $u$, we use $d(v,u)$ to denote the number of edges in the shortest path from $u$ to $v$.
Given a batch of targeted nodes $\mathcal{V}_{\mathcal{B}}$, for any node $u$, we define the distance between $u$ and  $\mathcal{V}_{\mathcal{B}}$ as $d(\mathcal{V}_{\mathcal{B}}, u) = \min(\{d(v,u)\}_{v \in \mathcal{V}_{\mathcal{B}}})$.
%From \autoref{eq:gnn}, we observe that 
After going deep into the computation paradigm of GNN models, we have the following observation.
\emph{Given the $k$\textsuperscript{th}-layer embedding, the receptive field of the next $(k+1)$\textsuperscript{th}-layer embedding become the $1$-hop neighborhood}. 
This observation motivate us to prune unnecessary nodes and edges from $\mathbf{A}_{\mathcal{B}}$.
Specifically, in the $k$\textsuperscript{th} layer, we prune every node $u$ with $d(\mathcal{V}_{\mathcal{B}}, u) > K-k+1$, as well as its associated edges, from $\mathbf{A}_{\mathcal{B}}$ to generate a pruned adjacent matrix $\mathbf{A}_{\mathcal{B}}^{(k)}$.
Therefore, \autoref{eq:gnn_batch} is revised as follows:
\begin{equation}\label{eq:gnn_batch_revised}
\mathbf{H}_{\mathcal{B}}^{(k+1)} = \Phi^{(k)}(\mathbf{H}_{\mathcal{B}}^{(k)}, \mathbf{A}_{\mathcal{B}}^{(k)}, \mathbf{E}_{\mathcal{B}}; \mathbf{W}_{\Phi}^{(k)}).
\end{equation}

Note that if we treat the adjacency matrix as a sparse tensor, only non-zero values are involved in model computation.
Essentially, the graph pruning strategy is to reduce the non-zero values in the adjacency matrix of each layer. 
%As we gradually prune nodes and edges durning different iterations. 
Therefore, it truly helps reduce unnecessary computations for most GNN algorithms. 
Moreover, each $\mathbf{A}_{\mathcal{B}}^{(k)}$ can be pre-computed in the subgraph vectorization phase. 
With the help of training pipeline strategy,  it takes nearly no extra time to perform graph pruning.
The right part of \autoref{fig:train_work_flow} gives a toy example to illustrate the graph pruning strategy \wrt one targeted node (\ie node A).

\textbf{Edge partitioning}. 
As shown in \autoref{eq:gnn_batch_revised}, the aggregator $\Phi^{(k)}$ is responsible to aggregate information for each node along its edges in the sparse adjacent matrix $\mathbf{A}_{\mathcal{B}}^{(k)}$. 
Several aggregation operators, such as sparse matrix multiplication, will be applied very frequently during the model computation phase, which makes the optimization of aggregation become very essential for GML system.
However, the conventional deep learning frameworks (\eg TensorFlow, PyTorch) seldom address this issue since they are not specially designed for GML technique. 

Tackling this problem, we propose an edge partitioning strategy to perform graph aggregation in parallel. 
The key insight is that a node only aggregates information along the edges pointing at it.
If all edges with the same destination node can be handle with the same thread, the multi-thread aggregation could be very efficient since there will be no conflicts between any two threads.
To achieve this goal, we partition the sparse adjacent matrix into $t$ parts and ensure that the edges with the same destination node (\ie the entries in the same row) fall in the same partition.
The edge partitioning strategy is illustrated in the top of the middle part of  \autoref{fig:train_work_flow}. 
After edge partitioning, each partition will be handle with a thread to perform aggregation independently. 
On one hand, the number of nodes in a batch of training examples is usually much larger than the number of threads.
On the other hand, the amount of neighbors for each node (\ie the number of non-zero entries in each row) will not be too large after applying sampling in \preprocessing.
Therefore, the multi-thread aggregation can achieve load balancing thus gains a significant speedup when training GNN models.

\begin{figure}[htbp]
\centering
\includegraphics[width=1.0 \linewidth]{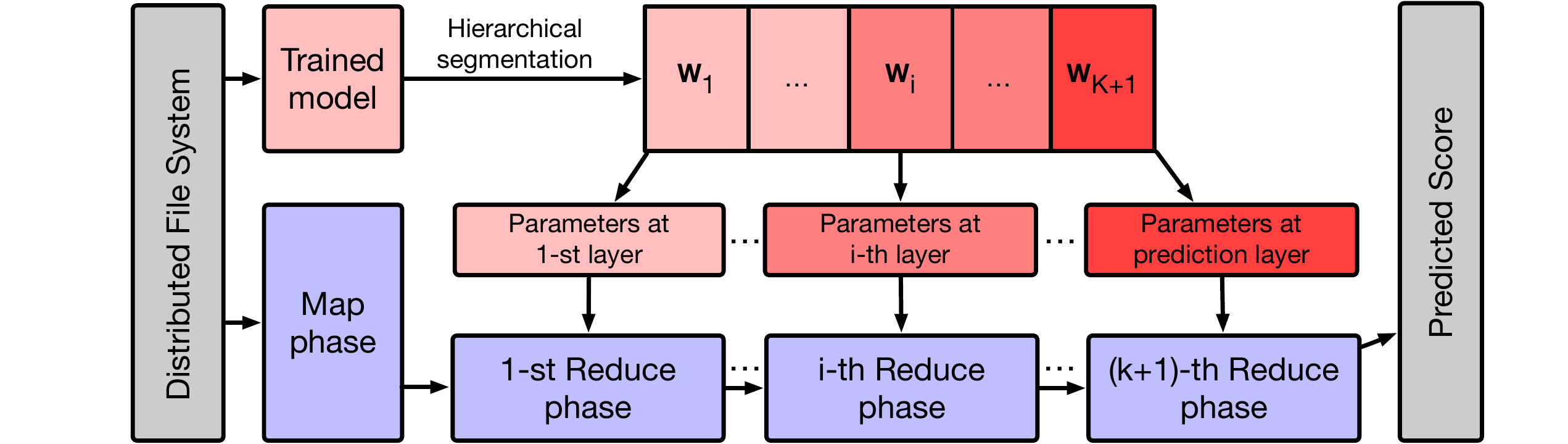}%{Figure9_graphinfer_tuotu_0304.eps}
\caption{The pipeline of GraphInfer. }
\label{fig:tuotu}
\end{figure}

\subsection{GraphInfer: distributed framework for GNN model inference}
Performing GNN model inference over the industrial-scale graphs could be an intractable problem.
%, and few solutions in literature has paid attention on this problem in real industry scenarios.
On one hand, the data scale and use frequency of inference tasks could be quite higher than that of training tasks in industrial scenarios, which require a well-designed inference framework to boost the efficiency of inference tasks.
On the other hand, since different $k$-hop neighborhoods described by \emph{GraphFeatures} could overlap with each other, directly performing inference on \emph{GraphFeatures} could lead to massive repetitions of embedding inference thus becomes time-consuming.

Hence, we develop \infer, a distributed framework for GNN model 
inference over huge graphs by following the message passing scheme.
%\infer first perform \emph{hierarchical model segmentation} to split a GNN model into different segments in terms of the model hierarchy. 
%It's common that the amount of data records in inference phase can be several orders of magnitude more than that in training phase, let alone inferring over the entire graph with billions of nodes and hundreds of billion edges.
%Since different $k$-hop neighborhoods may overlap with each other, the intermediate embeddings of nodes will be repeatedly computed, which makes the inference phase quite time-consuming.
%which leads to a high IO overhead and many repeated computations.
%To efficiently perform the inference task in industry scenarios, we build an MapReduce pipeline to mitigate this problem.
%Hence, we leverage a model-split technique and develop a MapReduce pipeline to solve this problem.
We first perform hierarchical    
model segmentation to split a well-trained $K$-layer GNN model into $K+1$ slices in terms of the model hierarchy.
Then, based on the message passing schema, we develop a MapReduce pipeline to infer with different 
slices in the order from lower layers to higher layers.
Specifically, the $k$\textsuperscript{th} \emph{Reduce} phase 
loads the $k$\textsuperscript{th} model slice, \emph{merges} the embeddings of last layer from in-edge neighbors to generate intermediate 
embeddings of the $k$\textsuperscript{th} layer and \emph{propagate} 
those intermediate embeddings via the out-edges to the destination 
nodes for the next \emph{Reduce} phase.
\autoref{fig:tuotu} describes the overall architecture 
of \infer, which can be summarized as follows:

%Model inference is also a challenging problem in industrial scenarios. On one hand, the scale of graphs in those situations may arrive at an unbelievable high level (as much as billons or even trillions of nodes and edges), while the number of node embeddings to be computed in inference procedure is usually much more than that in training procedure. On the other hand,  inference based on GraphFlat will lead to many repeated computations due to the overlap between different $k$-hop buckets (or $k$-hop subgraphs). Hence, inference by in the same way with that in training procedure may be not efficient enough in industrial scenarios. 

%To solve this problem, we design a MapReduce pipeline named TuoTu to perform the inference task.

%infer (`\emph{rule}') with a slice of the model and    
%The chief idea of TuoTu is to  use a MapReduce pipeline with $k$ reduce procedures to simulate the forward propagation of a GNN model with $k$ layers. As shown in Fig.~\ref{fig:tuotu}, at first, a $k$-hop GNN model will be split into $k$ slices and the $i$-th slice represents parameters of the $i$-th iteration in forward propagation procedure. Then, the $i$-th reducer takes the $i$-th slice to compute the $i$-th intermediate embeddings. The total procedure can be concluded as follows:
\begin{enumerate}
\item \textbf{Hierarchical model segmentation}. A $K$-layer GNN model is split into $K+1$ slices in terms of the model hierarchy. 
Specifically, the $k$\textsuperscript{th} slice ($k \leq K$) consists of all parameters of the $k$\textsuperscript{th} GNN layer, while the $K+1$\textsuperscript{th} slice consists of all parameters of the final prediction model.
\item \textbf{Map}. 
Similar to \preprocessing, the \emph{Map} phase here also runs only once at the beginning of the pipeline. 
For a certain node, the \emph{Map} phase also generate three kinds of information,  \ie the self information, the in-edge information and the out-edge information, respectively. 
Then, the node id is set as the ``shuffle key'' and the various information as the ``value'' for the following \emph{Reduce} phase.
\item \textbf{Reduce}. 
The \emph{Reduce} phase runs $K+1$ times in which the former $K$ rounds is to generate the $K$\textsuperscript{th}-layer node embedding while the last round to perform the final prediction. 
For the former $K$ rounds, a reducer acts similar to that in \preprocessing. 
In the merging stage, instead of generating $k$-hop neighborhood, the reducer here loads its model slice to infer the node embedding based on the self information and in-edge information, and set the result as the new self information. 
Note that in the $K$\textsuperscript{th} round, the reducer infers the $K$\textsuperscript{th}-layer node embedding and only need to output it rather than all of the three information to the last \emph{Reduce} phase.
The last \emph{Reduce} phase is responsible to infer the final predicted score and output it as the inference result. 
%In each of the former $K$ round, a reducer first load the $k$\textsuperscript{th} slice of the trained GNN model and collects all \emph{values} with the same \emph{shuffle key}.
%Then it infers the $k$\textsuperscript{th}-layer node embedding of a node with \autoref{eq:gnn}, based on the self information and in-edge information, and set it as the node's new self information. 
%Next, the new self information is propagated along out-edges to construct new in-edge information \wrt the destination nodes pointed by out-edges. 
%All of the out-edge information remain unchanged for the next \emph{Reduce} phase. 
%At last, the reducer outputs the new data records, with the node ids and the updated information as the new \emph{shuffle key} and \emph{value} respectively.
%For $i$-th reduce stage, perform the inference phase in $i$-th layer via in-edges and broadcast $i$-th intermediate embeddings of node $v_i$ through out-edges.
\end{enumerate}

There is no repetitions of embedding inference in the above pipeline, which reduces the time cost in a great extent. 
Moreover, the pruning strategy similar to that in \training also works in this pipeline in the case the inference task is performed over a part of the entire graph.
It's worthwhile to note that we also implement the sampling and indexing strategies, which are introduced in \autoref{sec:sampling},  in \infer to maintain the consistence of data processing with that in \preprocessing, which can provide unbiased inference with the model trained based on \preprocessing and \training.

%since the MapReduce pipeline will shuffle data records uniformly, there will be no bias if we apply the sampling strategy as stated in section.~\ref{sec:sampling} with that in building training set.
%The Map procedure takes data in the same format with that in the MapReduce pipeline introduced in section.~\ref{sec:graphflat}, while the last step stores all $k$-hop node embeddings into a distributed file system for downstream tasks. TuoTu is quite efficient as no repeated computation is involved. In addition, if the number of nodes in the inference set is far less than that in the whole graph, the pruning strategy also works in this MapReduce pipeline. 

\subsection{Demonstration}
%Figure \ref{fig:code} demonstrates the pseudo-code of GCN algorithms in our system. 

\autoref{fig:code} demonstrates how to use \system to perform data generation with \preprocessing, model training with \training, and inference with \infer. In addition, we also give an example on how to implement a simple GCN model.

For each module stated in \autoref{sec:training},   we provide a well-encapsulated interface respectively.
 \emph{GraphFlat} is to transform raw inputs into $k$-hop neighborhoods.
 User only need to chose a sampling strategy and prepare a node table together with an edge table, to generates $k$-hop neighborhoods \wrt their target nodes.
Those $k$-hop neighborhoods are the inputs of \training and are formulated as a set of triples $\mathcal{B}=\{<TargetedNodeId, Label, GraphFeature>\}$ as stated in \autoref{sec:training}.
Then, by feed \emph{GraphTrainer} a set of configurations like the model name, input, distributed training settings (the number of workers and parameter servers) and so on, a GNN model will be trained distributedly on the cluster. 
After that, \emph{GraphInfer} will load the well-trained model together with the inference data to perform the inference procedure.
In this way, developers only need to care about the implementation of the GNN model.

Here, we take GCN as an example to show how to develop GNN models in \system.
First, we should parse GraphFeature as  adjacent  matrix, node feature matrix and edge feature matrix (if needed) with the \emph{subgraph\_vectorize} function.
After that, we generate a list of adjacent matrix if enable the pruning strategy by calling the \emph{pruning} function.
Then, the $k$\textsuperscript{th} element in the \emph{adj\_list} (\ie a pruned adjacent matrix) together with intermediate embeddings generated by the former layer (or raw node features) will be fed to the $k$\textsuperscript{th} layer.
Note that, in each ``GCNLayer'', by calling the \emph{aggregator} function, information will be aggregated to target nodes from their direct neighbors according to \autoref{eq:gnn}.
Though these interfaces, a GNN model can be implemented quickly and there is no difference from coding for a single machine.

\begin{figure}[htbp]
\centering
\begin{minipage}[t]{0.9\linewidth}
\begin{lstlisting}[ numbers=none, 
         numberstyle=\tiny,keywordstyle=\color{blue!70},
         commentstyle=\color{red!50!green!50!blue!50},frame=single,
         rulesepcolor=\color{red!20!green!20!blue!20},basicstyle={\tiny\ttfamily}, breaklines=true, 
         morekeywords={GraphFlat, aggregator,subgraph_vectorize, pruning,GraphTrainer,edge_partition,GraphInfer}, tabsize=2]

###########GraphFlat###########
GraphFlat -n node_table -e edge_table -h hops -s sampling_strategy;

###########GraphTrainer###########
GraphTrainer -m model_name -i input -t train_strategy -c dist_configs;

###########GraphInfer###########
GraphInfer -m model -i input -c infer_configs;

###########Model File###########
class GCNModel:
	def __init__(self, targetID, GraphFeatue, label, ...)
		# get adj, node_feature and edge_feature from GraphFeatue
		adj, node_feature, edge_feature = subgraph_vectorize(GraphFeatue)
		....
		# pruning edges for different layers
		adj_list = pruning(adj) 
			
	def call(adj_list, node_feature, ...):
		# initial node_embedding with raw node_feature,  like: 
		# node_embedding = node_feature
		....
		# multi-layers
		for k in range(multi_layers):
			node_embedding = GCNlayer(adj_list[k], node_embedding)
			# other process like dropout
			...
		target_node_embedding = look_up(node_embedding, targetID)
		return target_node_embedding
		
...

class GCNLayer:
	def __init__(...)
	# configuration and init weights
	...
	def call(self, adj, node_embedding):
		
		# some preprocess
		...
		# aggregator with edge_partition
		node_embedding = aggregator(adj, node_embedding) 
	return node_embedding
\end{lstlisting}
\end{minipage}
\caption{A demo example of using \system.}
\label{fig:code}
\end{figure}

\section{Experiment}
In this section, we conduct extensive experiments to evaluate the proposed \system system.

\subsection{Experimental settings}

\subsubsection{Datasets.}
We employ three datasets in our experiments, including two popular public datasets (Cora\cite{b26}, PPI\cite{b27}) and an industrial-scale social graph  provided by Alipay\footnote{https://www.alipay.com/} (called UUG, \textbf{U}ser-\textbf{U}ser \textbf{G}raph).
%Datasets used in our experiments include two popular public datasets (Cora, PPI) from the academic literature and one large-scale graph (User-User Network) from real-world tasks in Alipay\footnote{https://www.alipay.com/}:
\begin{itemize}
\item \textbf{Cora.} Cora is a citation network with 2708 nodes and 5429 edges. Each node is associated with 1433-dimensional features and belongs to one of seven classes.
\item \textbf{PPI.} PPI is a protein-protein interaction dataset, which consists of 24 independent graphs with 56944 nodes and 818716 edges in total. 
Each node contains 50-dimensional features and belongs to several of 121 classes.
\item \textbf{UUG.} UUG consists of massive social relations collected from various scenarios of Alipay, in which nodes represent users and edges represent various kinds of interactions between users.
It contains as many as $6.23 \times 10^9$ nodes and $3.38 \times 10^{11}$ edges. 
Nodes are described with 656-dimensional features and alternatively belongs to two classes.
% We use 656-dimensional features to describe each node, which alternatively belongs to two classes. 
To our best knowledge, it is the largest attributed graph for GML tasks in literatures. 
\end{itemize}

Following the experimental settings in \cite{b1,b2,b3},  \emph{Cora} and \emph{PPI} are divided into three parts as the training, validation, and test set, respectively. 
For the \emph{UUG} dataset, $1.2 \times 10^8$ nodes out of $1.5 \times 10^8$ labeled nodes are set as the training set, while $5 \times 10^6$ and $1.5 \times 10^7$ are set as the validation and test set, respectively.
Those three sets are exclusive to each other. 
Details about those three datasets are summarized in Table.~\ref{tab:summary_datasets}.

\begin{table}[t]
\caption{Summary of datasets}
\begin{center}
\label{tab:summary_datasets}
\begin{tabular}{c||ccc}
\toprule
Indices     &Cora   &PPI        & UUG \\
\midrule
\#Nodes       &2708 &56944 (24 graphs)        &$6.23 \times 10^9$    \\
\#Edges         &5429 &818716           &$3.38 \times 10^{11}$   \\
\#Node feature  &1433      &50            &656   \\
\#Classes     &7            &121(multilabel)          &2    \\
\#Train set   &140          &44906 (20 graphs)        &$1.2 \times 10^8$  \\
\#Validation set &500 &6514 (2 graphs)    &$5 \times 10^6$ \\
\#Test set    &1000      &5524 (2 graphs)         &$1.5 \times 10^7$   \\
 \bottomrule
\end{tabular}
\end{center}
\end{table}

\subsubsection{Evaluation} We design experiments to compare our proposed system with two famous open-source GML systems, to demonstrate the effectiveness, efficiency, and scalability of our system.

\textbf{Compared Systems and GNN models.} Two famous GML systems in open-source community are used for comparison:
\begin{itemize}
\item \textbf{DGL}\cite{b18}. Deep Graph Library (DGL) is a Python package that interfaces between existing tensor-oriented frameworks (\eg PyTorch and MXNet) and the graph structured data. 
\item \textbf{PyG}\cite{b24}. PyTorch Geometric (PyG) is a library for deep learning on irregularly structured input data such as graphs, point clouds and manifolds, built upon PyTorch\cite{b25}.
\end{itemize}
For each of the two system, we evaluate three widely-used GNNs (\ie GCN, GAT and GraphSAGE) on two public datasets (\ie Cora and PPI) as stated above, respectively. 
In addition, the performance of those GNNs reported in their original literatures \cite{b1,b2,b3} are used as baselines. 

For a fair comparison, we tune hyper-parameters (\eg learning rate, dropout ratio, \etc.) for those GNNs by comprehensively referring to the details reported in \cite{b18,b24} together with official guidelines of DGL and PyG. 
For experiments on Cora and PPI, the embedding size is set to 16 and 64 respectively. 
All GNN Models are trained at a maximum of 200 epochs with Adam optimizer~\cite{b31}. 
We record average results after 10 runs for each experiment to mitigate variance.
Note that, when evaluating the training efficiency on public datasets, all systems are operated on an exclusive container (machine) with same CPUs (Intel Xeon E5-2682 v4@2.50GHz) in standalone mode.

Specially, for experiments on UUG, we deploy our system on a cluster in Ant Financial to verify the real performance of our proposal in industrial scenarios. 
Note that, the cluster used here is not exclusive.
Different tasks may be running on this cluster at the same time, which is common in industrial environment.
We analysis the convergence curve and speedup ratio by varying the number of workers to demonstrate the scalability of our system.
However, to our best effort, both DGL (v0.3.1) and PyG (v1.3.1) fail to operate on UUG dataset, since the distributed mode is not well supported in those systems and running with standalone mode will cause the OOM problem.
Therefore, the results of DGL and PyG on UUG dataset are not included.
%due to the OOM problem and the related result are not include. 

\textbf{Metrics.} We conduct experimental evaluation of the compared systems from several aspects.
First, we report the \emph{accuracy} and \emph{micro-F$_1$} score on Cora and PPI following the evaluation protocol in \cite{b2,b3}, to demonstrate the \emph{effectiveness} of GNN models trained with the compared systems.
Second, we report the average \emph{time-cost} per epoch in the training phase to demonstrate the \emph{training efficiency} of the compared systems.
Moreover, we train a node classification model with the UUG dataset, and do inference over the whole User-User Graph. By reporting the \emph{time-cost} of both the training and inference phases, we demonstrate the superior \emph{efficiency} of our proposal in the industrial scenario.
Last but not the least, we report the \emph{convergence curves} and the \emph{speedup ratio} of training on the the industrial-scale UUG dataset, to verify the \emph{scalability} of our system.
%perform a set of experiments on the large-scale User-User network, draw convergence curves, and report speedup ratio to verify the scalability of our system.

\subsection{Results and Analysis}
In this section, we present experimental results with associated analysis following the evaluation protocol stated above. 

\begin{table}[t]
\caption{Effectiveness of different GNNs trained with different systems}
\begin{center}
\label{tab:result_cora_ppi}
  \scalebox{0.80}{
\begin{tabular}{c|c|cccc}
\toprule
Datasets    &Methods    &Original Results     & PyG     &DGL    & \system   \\
\midrule
{Cora}        &GCN      &0.813            &0.818          &0.811    &0.811\\
(Accuracy)          &GraphSAGE    &$-$              &0.821        &0.818    &0.827\\
{}      &GAT            &0.830            &0.831        &0.828    &0.830\\
\midrule
{PPI}         &GCN              &$-$            &0.575          &0.561    &0.567\\
(micro-F1)      &GraphSAGE      &0.598            &0.632        &0.636    &0.635\\
      &GAT      &0.973        &0.983      &0.976    &0.977\\
\midrule
{UUG}         &GCN              &$-$            &$-$          &$-$    & 0.681\\
(AUC)     &GraphSAGE      &$-$            &$-$        &$-$    &0.708\\
      &GAT      &$-$        &$-$      &$-$    &0.867\\
 \bottomrule
\end{tabular} }
\end{center}
\end{table}

% \begin{figure*}[h]
%      \centering
%      \begin{subfigure}[b]{0.45\linewidth}
%          \centering
%          \includegraphics[width=\textwidth]{F1_batchsize_2}
%          \caption{Micro-F1}
%          \label{fig:f1_bucket}
%      \end{subfigure}
%       %\hfill
%      \begin{subfigure}[b]{0.45\linewidth}
%          \centering
%          \includegraphics[width=\textwidth]{loss_batchsize_2}
%          \caption{Loss curves}
%          \label{fig:loss_bucket}
%      \end{subfigure}
%         \caption{Micro-F1 and Loss curves on PPI with different batch sizes}
%         \label{fig:c_bias}
% \end{figure*}
\begin{table*}[t]
\caption{Time-cost(s) per epoch on PPI in Standalone mode}
\begin{center}
\label{tab:result_Times}
\begin{tabular}{c|ccc|ccc|ccc}
\toprule
{}              &{}     &GCN  &{}     &{}   &GraphSAGE  &{}   &{}   &GAT  &{}   \\
%\midrule
{}              &$1$-layer  &$2$-layer  &$3$-layer  &$1$-layer  &$2$-layer    &$3$-layer  &$1$-layer  &$2$-layer  &$3$-layer  \\
\midrule
PyG             &3.49 &6.43 &9.62 &4.47 &6.98   &10.15  &$44.29$    &$65.32$    &$85.21$\\
DGL             &1.09 &1.35 &1.62 &1.14 &1.39   &1.64 &16.14  &21.47  &26.03  \\
\midrule
{\system}\textsubscript{base}         &0.48 &2.75 &4.10 &0.46 &2.47   &3.94 &4.75 &25.72  &36.86\\
{\system}\textsubscript{+pruning}         &0.48 &1.93 &3.23 &0.46 &1.67   &2.99 &4.75 &13.88  &20.01\\
{\system}\textsubscript{+partition}   &0.42 &1.22 &1.60 &0.34 &0.97   &1.39 &4.63 &22.65  &33.45\\
{\system}\textsubscript{+pruning\&partition}  &\textbf{0.42}  &\textbf{1.13}  &\textbf{1.52}  &\textbf{0.34}  &\textbf{0.88}    &\textbf{1.35}  &\textbf{4.63}  &\textbf{13.73} &\textbf{18.63}\\
%{}       &GCN      &0.813            &0.818          &0.811    &0.811\\
%Cora           &GraphSage    &$-$              &0.821        &0.818    &0.827\\
%{}     &GAT            &0.830            &$-$        &0.828    &0.830\\
%\midrule
%{}         &GCN              &$-$            &0.575          &0.561    &0.567\\
%PPI      &GraphSage      &0.598            &0.632        &0.636    &0.635\\
%{}       &GAT      &0.973        &$-$      &0.976    &0.968\\
 \bottomrule
\end{tabular}
\end{center}
\end{table*}

\subsubsection{Evaluation on Public Datasets} We report the comparison of our system with DGL and PyG on two public datasets (i.e. Cora and PPI) with the following goals in mind:
\begin{itemize}
\item Compare the performance of some general GNN models in three system to verify the \emph{effectiveness} of  GNN models trained with \system.
\item Compare the time-cost of those GNN models in three systems to evaluate their \emph{efficiency}.
\end{itemize}

\begin{comment}

\begin{figure}[t]
     \centering
     \begin{subfigure}[b]{0.48\linewidth}
         \centering
         \includegraphics[width=\textwidth]{F1_batchsize_2}
         \caption{Micro-F1}
         \label{fig:f1_bucket}
     \end{subfigure}
      %\hfill
     \begin{subfigure}[b]{0.48\linewidth}
         \centering
         \includegraphics[width=\textwidth]{loss_batchsize_2}
         \caption{Loss curves}
         \label{fig:loss_bucket}
     \end{subfigure}
        \caption{Micro-F1 and Loss curves on PPI with different batch sizes}
        \label{fig:c_bias}
\end{figure}
\end{comment}

\textbf{Effectiveness.} \autoref{tab:result_cora_ppi} demonstrates the effectiveness over two public datasets and a industrial dataset (\ie accuracy in Cora, micro-F1 in PPI, and AUC in UUG) of GCN\cite{b1}, GAT\cite{b2} and GraphSAGE\cite{b3} implemented with three compared GML systems, respectively. 
Meanwhile, we also report the results of these GNN models presented in their original literatures \cite{b1,b2,b3} as baselines.  

Obviously, for these two public datasets, the performance of all three GNN models implemented and trained using \system is comparable to the models in PyG and DGL. 
%all these three algorithms in our subgraph-based system achieve comparable results with those in PyG and DGL. 
In most cases, the performance deviation of a GNN model is less than 0.01. 
Specially, for GraphSAGE on PPI, the performance of three compared systems are higher than the baseline, which is due to the difference in the propagation phase.
Specifically, when propagating the aggregated information of neighbors to the targeted node, the three systems use an ``add'' operator while the baseline use a ``concat'' operator.

Furthermore, those three GNN models in \system work well and achieve reasonable results on UUG. 
We get comparable results for GCN and GraphSAGE, but witness a significant improvement for GAT. 
The reason is that, the GAT model learns different weights for neighbors, which may play different roles (\ie friend, colleague and so on) \wrt their targeted node and have different influences on it.
Note that, to our best effort, we fail to deploy UUG on PyG and DGL due to the OOM problem, which also prove the scalability of the proposed \system.

\begin{table*}[tbp]
\caption{Inference efficiency on User-User Graph.}
\begin{center}
\label{tab:infer_efficiency}
\begin{tabular}{c|c|ccc}
\toprule
Methods     &Phase        &Time-cost (s)    &CPU-cost (core*min)    &Memory-cost (GB*min) \\
\midrule
{}        & \preprocessing    &13454            &436016         &654024 \\
Original          & Forward propagation  &5760             &93240          &1053150  \\
{}      &Total              &18214            &529256         &1707174  \\
\midrule
GraphInfer      & Total              &4423           &267764         &401646 \\

 \bottomrule
\end{tabular}
\end{center}
\end{table*}

\textbf{Efficiency}. 
Based on PPI dataset, we train different GNN models (\ie GCN, GraphSAGE and GAT) with different depth of layers  (\ie $1$-layer, $2$-layer and $3$-layer) on three compared GML systems.
\autoref{tab:result_Times} reports the average time-cost per epoch of all training tasks.
It also shows the results of our system with different optimization strategies stated in \autoref{sec:training} (\ie graph pruning and edge partitioning). 
%Numbers in Table~\ref{tab:result_Times} represents time-cost (second) per epoch while the \textbf{Bold} number is used to indicate the best results. 
Specifically, the subscript \textsubscript{Base} means training only with the pipeline strategy, while \textsubscript{+pruning}, \textsubscript{+partition}, and \textsubscript{+pruning\&partition} represent training with graph pruning strategy, edge partition strategy, and both of them, respectively. 
Note that our system is specially designed for industrial-scale graphs. 
Durning the training phase, data will be loaded from disks rather than from memory (like PyG and DGL), we treat {\system}\textsubscript{Base} as baselines for fairness.
%Note that, GraphFlat is specially designed for industrial-scale graphs and data will be load from disks durning training phase, 
%we treat the \emph{Base} model as the baseline 
%Note that, with the pipeline strategy, 

Though our system is designed for distributedly training GNN models over industrial-scale graphs, it also demonstrates a gifted speed on CPUs in standalone mode.
%Those algorithms in our system (with pruning strategy and optimized aggregator) are consistently faster than those in DGL and PyG in terms of operating on CPUs in the standalone mode. 
Generally, in the training phase, our system achieve a \textbf{5}$\times$ $\sim$ \textbf{13}$\times$ speedup compared with PyG,  and a \textbf{1.2}$\times$ $\sim$ \textbf{3.5}$\times$ speedup compared with DGL.
For all three GNN models at different depths, the performance of our system is superior to the other two systems to varying degrees.
Specially, compared to PyG, our system achieves the greatest improvement, \ie a \textbf{7}$\times$ $\sim$ \textbf{13}$\times$ speedup, in the training of GraphSAGE. 
Compared to DGL, when training $1$-layer GNN models, our system gains more significant improvement, \ie a \textbf{2.5}$\times$ $\sim$ \textbf{3.5}$\times$ speedup.

%For GCN, our system gains \textbf{5}$\times$ $\sim$ \textbf{8}$\times$ speedup compared with PyG, while for GraphSAGE, our system is \textbf{7}$\times$ $\sim$ \textbf{13}$\times$ faster. 
%Meanwhile, our system also outperforms DGL in most situations. 
%For $1$-hop graph embedding models, our system achieves about \textbf{2}$\times$ $\sim$ \textbf{4}$\times$ speedup over DGL. 
%For models with more hops, our system also gains better or at least comparable performance compared with DGL. 
%For example, our system reduces the time-cost from 26.03s to 18.63s when learning a $3$-hop GAT model. 
%%% to be modified
%Note that, PyG crashes when performing GAT model and results in related columns are null in this Table.

Moreover, we further verify the superiority of the proposed optimization strategies, \ie graph pruning and edge partitioning, in \autoref{tab:result_Times}.
%the proposed strategies in section~\ref{sec:model_training} are proved to be effective in those experiments. 
%In Table~\ref{tab:result_Times}, we illustrate the comparison of models with strategies respectively. 
The observations from these results can be summarized as follows.
First, either the graph pruning strategy or the edge partitioning strategy works consistently well on different GNN models, which is proved by comparing the result of {\system}\textsubscript{+pruning} or {\system}\textsubscript{+patition} to that of {\system}\textsubscript{+base}.
Furthermore, when comparing the result of {\system}\textsubscript{+pruning} or {\system}\textsubscript{+patition} to that of {\system}\textsubscript{+pruning\&partition}, we observe that a greater improvement is achieved by combining these two optimization strategies together.
%Models with these two strategies gains \textbf{2.7}$\times$, \textbf{2.9}$\times$, and \textbf{2.0}$\times$ speedup for GCN, GraphSage, and GAT compared with the base model in terms of learning $3$-hop embeddings respectively. 
Second, these two strategies individually lead to different results in different situations. 
The edge partitioning strategy achieves better speedup ratio when applied in GCN and GraphSAGE than in GAT, while the graph pruning strategy doesn't work in training $1$-layer GNN model but demonstrates its power when training deeper GNN models.

These observations are caused by different insights behind the two strategies.
The graph pruning strategy aims to mitigating unnecessary computations by reducing edges that won't be used to propagate information to target nodes. 
The edge partitioning strategy achieves information aggregation among neighbors in an efficient parallel way.
On one hand, since these two strategies optimize some key steps of training GNN models, their advantages benefit the training of GNN models in general. 
On the other hand, there also exists some limits.
For example, if we train a $1$-layer GNN model, it's reasonable that the pruning strategy won't work, as every edge plays a role in propagating information to target nodes and there's no unnecessary computations.
Moreover, if a model consists of more dense computation (like computing attentions) than aggregating information along edges, the effect of these strategies will be weakened, since the dense computation takes the most part of the total time-cost.

\subsubsection{Evaluation on Industrial Dataset} 
We implement the proposed system using MapReduce and parameter server framework, and deploy it on a CPU cluster consisting of more than one thousand machines (Each machine is powered by a 32-core CPU with 64G memory and 200G HDD).
Then, we conduct experiments on the industrial dataset, \ie UUG dataset, to demonstrate the scalability and efficiency of the proposed system in industrial scenarios.
%We design a set of experiments on a large-scale graph from Alipay (User-User Network) to evaluate our system in terms of scalability and inference efficiency in real industrial scenarios.

\begin{figure}[htbp]
  \centering
  %\subfigure[Convergence]{
  \begin{minipage}[t]{0.49\linewidth}
    \centering
    \includegraphics[width=0.99\linewidth]{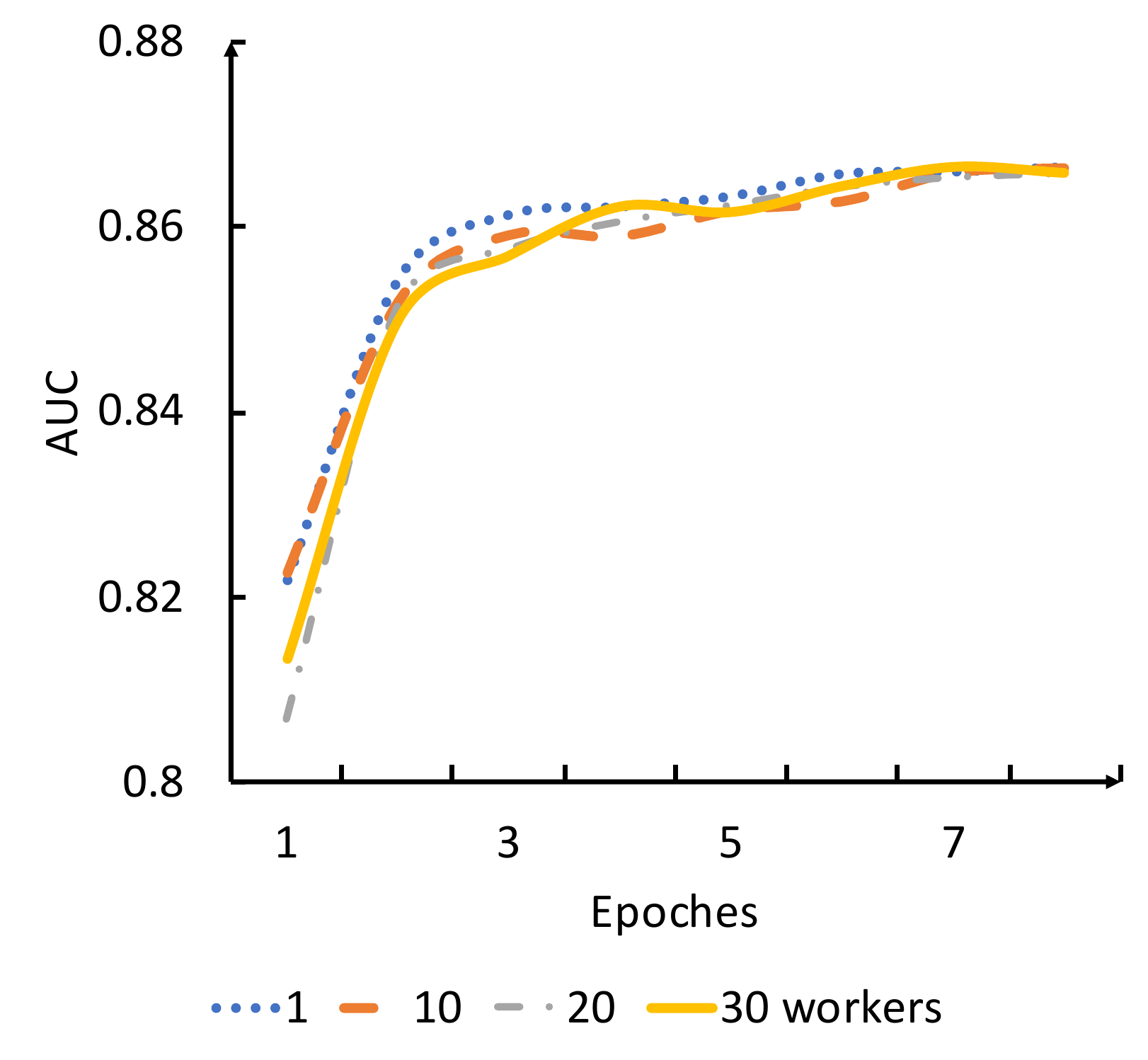}
    \caption{Convergence}
    \label{fig:convergence_worker_num}
  \end{minipage}
  %}
  %\subfigure[Speedup]{
  \begin{minipage}[t]{0.49\linewidth}
    \centering
    \includegraphics[width=0.99 \linewidth]{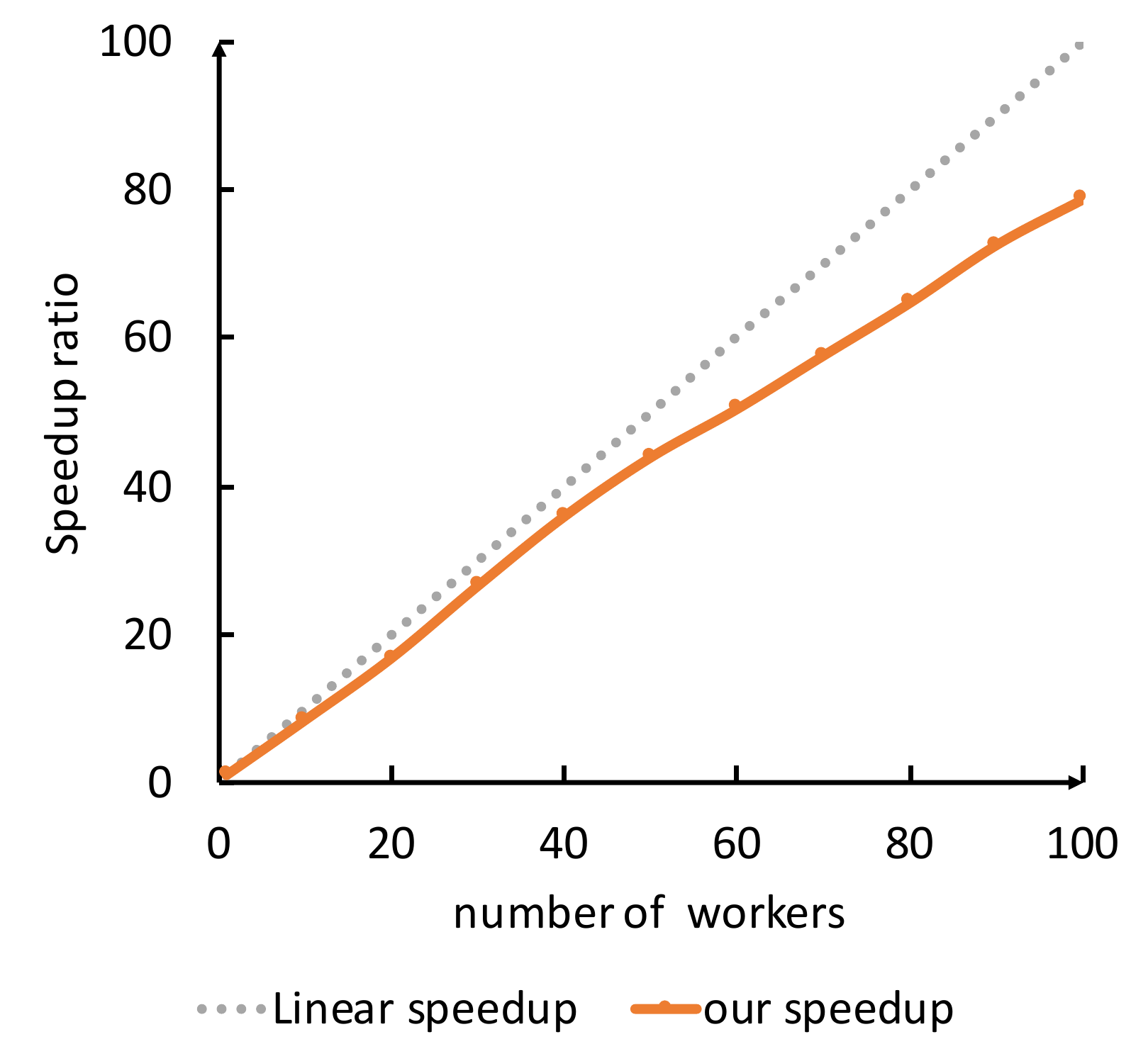}
    \caption{Speedup}
    \label{fig:speedup_worker}
  \end{minipage}
  %}
%\caption{ pics}
\end{figure}

\textbf{Industrial training}. 
Scalability is one of the most important criterions for industrial GML systems. 
In this subsection, we focus on evaluating the training scalability of \system on two aspects, \ie \emph{convergence} and \emph{speedup}. 
To do that, we train a GAT model on the industrial UUG dataset with different number of workers and report the results of convergence and speedup in \autoref{fig:convergence_worker_num} and \autoref{fig:speedup_worker}, respectively.

\emph{Convergence}.
\autoref{fig:convergence_worker_num} demonstrates the training scalability of our system in terms of convergence.
Its y-axis denotes the AUC of GNN model, while the x-axis denotes the number of training epochs. 
In general, our system eventually converge to the same level of AUC regardless the number of training workers.
As shown in \autoref{fig:convergence_worker_num}, though more training epochs are required in the distributed mode, the convergence curves finally reach the same level of AUC as that trained with a single work.
%models with multi-workers converge with more training epochs than the model using just a single worker, but they finally arrive at a same place. 
Hence, the model effectiveness is guaranteed under distributed training, which verifies that our system have the ability to scale to industrial graphs without considering convergence.
%we can distributedly train a model with our system when scaling to large-scale graphs without any concerns in convergence.

\emph{Speedup}.  
We also demonstrate the training scalability in terms of speedup ratio.
As shown in \autoref{fig:speedup_worker}, our system achieves a near-linear speedup with slope ratio about 0.8, which means that if you double the number of training workers, you will get \textbf{1.8}$\times$ faster. 
In the experiment, we scale the number of training workers form 1 to 100 with 10 intervals.
As a result, our system achieve a constantly high speedup and the slope ratio hardly decreases. 
For example, when the number of training workers reaches 100, we have \textbf{78}$\times$ faster, which is only slightly lower than the expect value 80. 
Note that, all these experiments are conducted on a cluster in the real production environment.
There may exist different tasks operating on a same physical machine. 
The overhead in network communication may slightly increase as the number number of training workers increases, causing perturbations in the slope ratio of the speedup curve. 
That again proves the robustness of our system in the industrial scenario. 

% It's worthing noting that, it only takes about 10 hours to train a two-layer GAT model on UUG until convergence, which is remarkable for industrial applications.
% In out experiment, the GraphFlat phase takes about 3.7 hours with 1000 workers to prepare GraphFeature, while the training phase takes about 10 hours with only 100 workers.
% Furthermore, in training phase, the training task only need 5.5 GB memory for each workers (550 GB in total), which is far less than memory cost for the total graph (2.92 TB).

It's worthing noting that, it only takes about 14 hours to train a 2-layer GAT model on UUG until it converges to a stable state.
Specifically, in out experiment, the \preprocessing takes about 3.7 hours with 1000 workers to generate GraphFeature, while the \training takes about 10 hours with only 100 workers on the CPU clusters to train a GAT model.
The total pipeline can be finished in 14 hours, which is remarkable for industrial applications.
Furthermore, during the training phase, the training task only need 5.5 GB memory for each workers (550 GB in total), which is far less than the memory cost for storing the entire graph (35.5 TB).

In summary, thanks to its ingenious architectural design, the proposed \system meets the industrial scalability requirements for training GNN models over industrial graphs. 
%In all, those results prove that our system truly gains a high near-linear speedup in real industrial scenarios.
%average time-cost per epoch. 
%As shown in Fig.~\ref{fig:speedup_worker}, our system achieves a near-linear speedup when using more machines, which means that, with the increasing of machines that we used for training, the time-cost during training procedure will get a linearly reduction. 
%For example, when we scale the number of machines from 1 to 100 with 10 intervals, our system is \textbf{8.8}$\times$, \textbf{15.1}$\times$, \textbf{22.2}$\times$ faster. 
%Note that, all those experiments are operated on clusters in a production environment, in which, different tasks may operate on a same machine. 
%Hence, those results prove that our system truly gains a high near-linear speedup in real industrial scenarios.

%\begin{figure}[htbp]
%\centering
%\includegraphics[width=0.98 \linewidth]{convergence_vs_workers_0210}
%\caption{Convergence curves of models with different number of workers. }
%\label{fig:convergence_worker_num}
%\end{figure}
%
%\begin{figure}[htbp]
%\centering
%\includegraphics[width=0.98 \linewidth]{speedup_0210}
%\caption{Speedup ratio vs. different number of workers. }
%\label{fig:speedup_worker}
%\end{figure}

\textbf{Industrial inference}. We evaluate the efficiency of \infer over the entire User-User Graph, which consists of $6.23 \times 10^9$ nodes and $3.38 \times 10^{11}$ edges.
In \autoref{tab:infer_efficiency}, we report the time and resource consumed by such an inference task.
Since no GML system can handle such a large scale graph, we compare \infer with the original inference module based on \emph{GraphFeature}. 
Note that, all these experiments are operated with the same concurrency, \ie 1000 workers.

From \autoref{tab:infer_efficiency}, we can observe that \infer consistently outperforms the original inference module in both time-cost and resource-cost. 
\infer takes about 1.2 hour to infer the predicted scores of 6.23 billion nodes with a 2-layer GAT model generating 8-dimensional embedding, which is just about $\frac{1}{4}$ of the time spent by the original inference module. 
Moreover, \infer also saves $50\%$ of CPU-cost and $76\%$ of memory-cost, respectively.
Compared with the original inference module based on \emph{GraphFeature}, \infer avoids repeated computing by employing the message passing scheme, which is the reason why it outperforms the original inference module.

\section{Conclusion}

In this paper, we present \system, an integrated system designed
for industrial-scale graph learning tasks. Our system design follows
the message passing scheme underlying the computation of GNNs, where
we simply merge values from in-edge neighbors and propagate merged
values to out-edge neighbors. With this programming principle, we
design to implement the construction of $k$-hop neighborhood,
an information-complete subgraph for each node, and the inference
in MapReduce. In addition, the $k$-hop neighborhood ensures the independency
among nodes in the graph, thus makes us simply train the model with 
parameter servers. \system maximally utilizes the calculation
of each embedding at inference, while optimizes the training from 
model level to operator level. As a result, \system successfully
achieves a nearly linear speedup in training with 100 workers. \system can finish
the training of a 2-layer graph attention network on a graph with billions of nodes and hundred billions of edges in 14 hours, and complete the inference
in only 1.2 hour. We have all these achivements based only on mature   
infrastructures such as parameter server and MapReduce.

\balance

%ACKNOWLEDGMENTS are optional

%\section{Acknowledgments}
%This section is optional; it is a location for you
%to acknowledge grants, funding, editing assistance and
%what have you.  In the present case, for example, the
%authors would like to thank Gerald Murray of ACM for
%his help in codifying this \textit{Author's Guide}
%and the \textbf{.cls} and \textbf{.tex} files that it describes.

\bibliographystyle{abbrv}
\bibliography{vldb_bib}  % vldb_sample.bib is the name of the Bibliography in this case

%\subsection{References}
%Generated by bibtex from your ~.bib file.  Run latex,
%then bibtex, then latex twice (to resolve references).

%APPENDIX is optional.
% ****************** APPENDIX **************************************
% Example of an appendix; typically would start on a new page
%pagebreak

%\begin{appendix}
%You can use an appendix for optional proofs or details of your evaluation which are not absolutely necessary to the core understanding of your paper. 
%
%\section{Final Thoughts on Good Layout}
%Please use readable font sizes in the figures and graphs. Avoid tempering with the correct border values, and the spacing (and format) of both text and captions of the PVLDB format (e.g. captions are bold).
%
%At the end, please check for an overall pleasant layout, e.g. by ensuring a readable and logical positioning of any floating figures and tables. Please also check for any line overflows, which are only allowed in extraordinary circumstances (such as wide formulas or URLs where a line wrap would be counterintuitive).
%
%Use the \texttt{balance} package together with a \texttt{\char'134 balance} command at the end of your document to ensure that the last page has balanced (i.e. same length) columns.
%
%\end{appendix}

\end{document}